\documentstyle[12pt,epsf]{article}
\begin{document}
\baselineskip 14pt plus 2pt minus 2pt
\newcommand{\beq}{\begin{equation}}
\newcommand{\eeq}{\end{equation}}
\newcommand{\beqa}{\begin{eqnarray}}
\newcommand{\eeqa}{\end{eqnarray}}
\newcommand{\dfrac}{\displaystyle \frac}
\renewcommand{\thefootnote}{\#\arabic{footnote}}
\newcommand{\ve}{\varepsilon}
\newcommand{\krig}[1]{\stackrel{\circ}{#1}}
\newcommand{\barr}[1]{\not\mathrel #1}
\newcommand{\vs}{\vspace{-0.25cm}}
\renewcommand{\arraystretch}{1.3}

\begin{titlepage}

{\small

\noindent Accepted for publication in Eur. Phys. J. A
\hfill FZJ-IKP(TH)-1998-07

\hfill LPT-98-03}

\vspace{1.0cm}

\begin{center}

{\Large  \bf {
Novel approach to pion and eta  production \\ [0.4em]
in proton-proton collisions near threshold\footnote{Work supported by DFG under
contract no. 445 AUS--113/3/0.}
}}

\vspace{1.cm}

{\large V. Bernard$^{\ddag}$,
N. Kaiser$^{\diamond}$,
Ulf-G. Mei\ss ner$^{\dag}$}

\vspace{0.7cm}

$^{\ddag}$Universit\'e Louis Pasteur, Laboratoire de Physique
Th\'eorique\\ 3-5, rue de l'Universit\'e, F--67084 Strasbourg, France\\
{\it email: bernard@lpt1.u-strasbg.fr} \\

\vspace{0.3cm}
$^{\diamond}$Technische Universit\"at M\"unchen, Physik Department T39\\
James-Franck-Stra{\ss}e, D--85747 Garching, Germany\\
{\it email: nkaiser@physik.tu-muenchen.de}\\

\vspace{0.3cm}
$^{\dag}$Forschungszentrum J\"ulich, Institut f\"ur Kernphysik (Theorie)\\
D--52425 J\"ulich, Germany\\
{\it email: Ulf-G.Meissner@fz-juelich.de} \\

\end{center}

\vspace{0.9cm}

\begin{abstract}
\noindent We evaluate the threshold matrix--element for the reaction
$pp \to pp\pi^0$ in a fully relativistic Feynman diagrammatic approach.  We
employ a simple  effective range approximation to take care of the S--wave $pp$
final--state interaction. The experimental value for the threshold amplitude
${\cal A} = (2.7 - i\,0.3)$ fm$^4$ can be reproduced by contributions from tree
level chiral (long--range) pion exchange and short--range effects related to
vector meson exchanges, with $\omega$ exchange giving the largest individual
contribution. Pion loop effects appear to be small. We stress that
the commonly used heavy baryon formalism is not applicable in the NN--system
above the pion production threshold due to the large external momentum, $|\vec
p\,| \simeq \sqrt {Mm_\pi}$, with $M$ and $m_\pi$ the nucleon and the
pion mass, respectively. We furthermore investigate the reaction $pp\to p
n \pi^+$ near threshold within the same approach. We extract from the data the
triplet threshold amplitude as ${\cal B}= (2.8 -i\,1.5)$ fm$^4$. Its real
part can be well understood from (relativistic) tree level meson--exchange 
diagrams. In addition, we investigate the process $pp \to pp \eta$ near
threshold. We use a simple factorization ansatz for the  $pp\eta$ final--state
interaction and extract from the data the modulus of the threshold amplitude,
$|{\cal C}| = 1.32\,$fm$^4$.  With $g_{\eta N}=5.3$, this value can be
reproduced by (relativistic) tree level meson--exchange diagrams and
$\eta$--rescattering, whose strength is fixed by the $\eta N$ scattering
length. We also comment on the recent near threshold data for $\eta
'$--production. 
\end{abstract}


\vspace{2cm}

\vfill

\end{titlepage}


\section{Introduction and summary}
\label{sec:intro}

With the advent of proton cooler synchrotrons at Bloomington, J\"ulich and
Uppsala, high precision data for the processes $pp \to pp\pi^0$, $pp\to
d\pi^+$, $pp\to pn\pi^+$ and $pp \to pp \eta$ in the threshold
region~\cite{meyer,bondar,who,pipn,calen} have
become available. The first data on neutral pion production were a big
surprise because the experimental cross sections turned out to be a factor of
five larger than the theoretical predictions based on direct pion production
and neutral pion rescattering fixed from on-shell $\pi N$ data. Nevertheless,
the energy--dependence of the total cross section $\sigma_{\rm tot}$
from threshold to excess energies of about 50~MeV
is completely given by the strong S--wave $pp$ final--state interaction, see
e.g. ref.\cite{MuS}. Subsequently, it was argued that heavy--meson exchanges
might be able to remove this discrepancy~\cite{LR,chuck}. On the other hand it
was shown~\cite{oset,unsers} that the (model--dependent) off--shell
behavior of the full $\pi N$ T--matrix also can enhance the cross section
considerably. Another avenue to eventually understand the near threshold cross
section is offered by calculations within the framework of tree--level heavy
baryon chiral perturbation theory (HBChPT) including dimension two
operators~\cite{pmmmk,bira1,bira2,lee}. In these papers, HBChPT has been used
to constrain the long--range $\pi^0$-exchange contributions. However, it
was found that the calculations  for $pp \to pp\pi^0$ performed so far lead to
a marked difference in the role of the so--called $\pi^0$ rescattering
contribution, which interferes constructively with the direct production in the
J\"ulich model and destructively in the HBChPT framework, respectively. In a
recent paper, it was argued that the treatment underlying the isoscalar
pion--nucleon scattering amplitude and the related transition operator
for the process $NN \to NN\pi$ (where $N$ denotes the nucleon)
 in the HBChPT framework is not yet sufficiently
accurate and thus the resulting rescattering contribution should be considered
as an artifact of this approximation~\cite{hhhms}. It was also stressed that
the process $pp\to d\pi^+$ is more sensitive to the long--range (chiral) pion
exchange. One--loop graphs have been considered in~\cite{isra}.

\medskip

However, one important feature of this reaction should be kept in mind, namely
the large momentum mismatch between the initial and the final nucleon--nucleon
state: the initial center--of--mass momentum is given by $|\vec p \, |^2 =
Mm_\pi + {m_\pi^2}/{4}$, with $m_\pi \,(M)$ the neutral pion (proton) mass,
whereas close to threshold the final one is compatible with zero. This
large difference leads to momentum transfers equal to $|\vec p\,|$ and higher.
In particular, both the squared invariant momentum transfer and the kinetic
energy of the incoming protons are $-Mm_\pi$ and $m_\pi/2$, respectively, and
therefore the chiral counting in the original sense does not apply, see
e.g.~\cite{bira1}. We will actually demonstrate here that non-relativistic
approximations (i.e. the heavy baryon formalism or HBChPT) are not applicable
in the two-nucleon system above the pion production threshold. The simple
reason for that is the extreme kinematics with the external momentum
$|\vec p\,| \simeq \sqrt{Mm_\pi}$ diverging in the heavy nucleon limit $M\to
\infty$ (while this argument could be considered formal, it nicely pinpoints
the problems of the heavy baryon approach). The
HBChPT framework (which is very successful for low--energy single nucleon
processes) looses its systematic order--by--order low-energy
expansion when external momenta grow with the nucleon mass ($|\vec p\,| \sim
\sqrt{M}$). In order to avoid such problems (which are simply related to
kinematics) it seems mandatory to perform fully relativistic calculations. In
particular, we find that one of the low--energy constants $c_2$ entering the
chiral $\pi^0$--rescattering is enhanced in the relativistic calculation by a
factor of two as compared to the  non-relativistic HBChPT approach.
In this spirit we perform here a fully relativistic calculation of various tree
and one-loop diagrams contributing to $pp\to pp \pi^0$ and $pp \to pn \pi^+$ at
threshold. Our approach can also be extended in a straightforward
manner to $\eta$ and $\eta'$ production in $pp$--collisions. We
thus discuss here also the reaction $pp \to pp \eta$ up to laboratory excess
energies of about 100~MeV using a simple factorized form to account for the
final--state interactions in the $pp\eta$ three--body system.

\newpage

The pertinent results of this investigation can be summarized as follows:

\begin{enumerate}

\item[(i)] Assuming that the $pp$ final--state interaction is an on--shell
NN--process and using a simple effective range parametrization for the $^1S_0$
$pp$ phase shift, we can accurately fit the 40 data points of the total
cross section from threshold up to $T_{\rm lab} = 326$ MeV with a constant
(threshold) amplitude equal to  ${\cal A} = (2.7 - i\,0.3)$ fm$^4$.

\item[(ii)] The real part of this number can be well understood in terms of
chiral $\pi^0$ exchange (including chiral $\pi^0$ rescattering) and heavy meson
($\omega,\, \rho^0,\, \eta$) exchanges based on a relativistic Feynman diagram
calculation.

\item[(iii)] We have evaluated some classes of one-loop graphs and find that
they lead to small corrections of the order of a few percent.
Therefore chiral loops do not seem to play any significant role
in the processes $NN\to NN \pi$,  which are dominated by one--pion exchange
and short--range physics (with the notable
exception of $pp \to d\pi^+$, where $d$ denotes the deuteron).

\item[(iv)] Both the long range $\pi^0$ exchange and the short range vector
meson exchange lead to contributions to the threshold amplitude ${\cal A}$
which do not vanish in the chiral limit $m_\pi \to 0$. There is no
chiral suppression of the reaction $pp\to pp \pi^0$ compared to other $NN\pi$
channels.  In all cases the respective threshold amplitudes are non-zero (and
finite) in the chiral limit. This is in contrast to the widespread believe
that $pp \to pp \pi^0$ is suppressed for reasons of chiral symmetry.

\item[(v)] Within the same approach, we have investigated the threshold
behavior of the process $pp \to pn\pi^+$. It is given in terms of ${\cal A}$
and the triplet threshold amplitude ${\cal B}$ with the empirical value
${\cal B}=(2.8 -i\,1.5)$ fm$^4$. The corresponding real part Re$\,{\cal B}$ is
well reproduced by chiral one--pion exchange and short--range vector meson
physics. The empirical value of ${\cal B}$ has however a sizeable imaginary
part, which naturally can not be explained by tree graphs. This channel
deserves some further study. The small and large imaginary parts of ${\cal A}$
and ${\cal B}$, respectively, are related to the weak and strong initial state
interaction in the $^3P_0$ and $^3P_1$ entrance channel, respectively.

\item[(vi)] Finally, we have investigated the process $pp \to pp \eta$ near
threshold. We treat the final state--interaction in the $pp$ and $\eta p$
subsystems using a factorization ansatz and effective range approximations. The
empirical value of the modulus of the threshold amplitude, $|{\cal
  C}|= 1.32$ fm$^4$, is reproduced with $g_{\eta N}=5.3$ by tree level meson
exchange graphs and $\eta$--rescattering. The strength of the latter is fixed
from the (real part of the) $\eta N$ scattering length.

\item[(vii)] Altogether, it is quite surprising that the total cross sections 
for pion and  eta production in $pp$--collisions near threshold can be 
explained so simply in terms of final--state interaction (using effective 
range approximations) and certain well-known (relativistic) meson--exchange
diagrams for the respective threshold amplitudes. However, such an approach 
requires further improvements and tests. First, one should  include
systematically initial--state and final--state NN--interactions. Secondly, in
order to perform a detailed test of the underlying meson production mechanism
one has to consider more exclusive observables like angular distributions of
differential cross sections and asymmetries generated by polarized proton beams
and targets. We hope to report on these topics in the near future.  
\end{enumerate}

\medskip


\section{Threshold kinematics and final--state interaction}

\subsection{Kinematics}

\begin{figure}[htb]
   \epsfysize=4.5cm
   \centerline{\epsffile{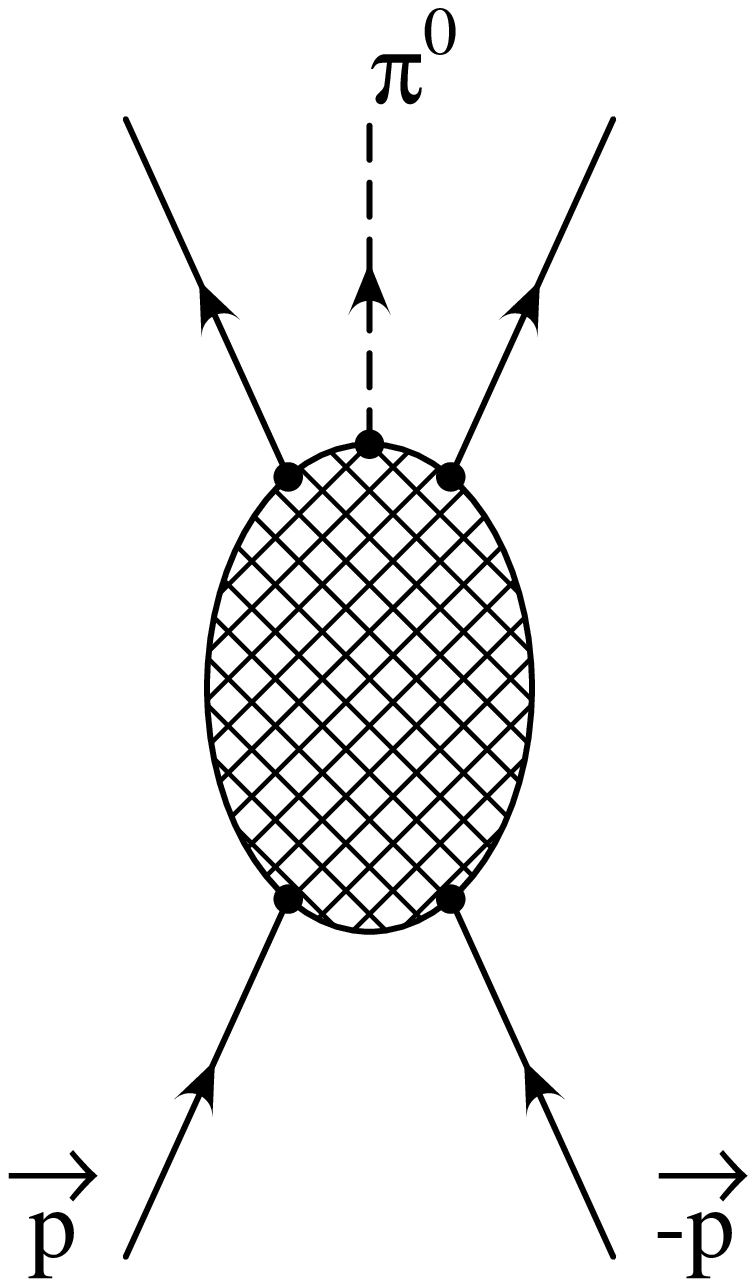}}
   \medskip
   \centerline{\parbox{13cm}{\caption{\label{fig1}
The process $pp \to pp \pi^0$ in the center--of--mass system.
  }}}
\end{figure}

We consider the reaction $p_1 (\vec{p}\,) + p_2 (-\vec{p}\,) \to p+p+\pi^0$ in
the center--of--mass (cm) frame at threshold (see fig.~\ref{fig1}). The
invariant T--matrix can be expressed in terms of one complex--valued (constant)
amplitude, which we denote by ${\cal A}$, as
\beq\label{T}
{\rm T}^{\rm cm}_{\rm th} (pp\to pp\pi^0) = {\cal A} \, (i\,
\vec{\sigma}_1 - i \,\vec{\sigma}_2 + \vec{\sigma}_1 \times
\vec{\sigma}_2 ) \cdot \vec{p} \quad .
\eeq
The $\vec{\sigma}_{1,2}$ are the spin--matrices of the two protons. The value
of the proton cm momentum to produce a neutral pion at rest is given by
\beq |\vec{p}\,| = \sqrt{m_\pi (M + m_\pi/4)} = 362.2~{\rm MeV}~,
\eeq
with $M = 938.27$~MeV the proton and $m_\pi = 134.97$~MeV the neutral pion
mass, respectively. Obviously, $|\vec{p}\,|$ vanishes in the chiral limit of
zero pion mass. Therefore the soft--pion theorem which requires a vanishing
threshold T--matrix in the chiral limit $m_\pi=0$ is trivially fulfilled (as
long as ${\cal A}$ does not become singular)~\cite{SSY}.
We remark that similar features
occur for the reaction $\pi N \to \pi \pi N$ (see ref.\cite{bkmppn1}).
All dynamical information is encoded in the threshold amplitude
${\cal A}$ of dimension [length$^4$]. In the threshold region, the wave
function of the final di--proton system as well as that of the neutral pion are
dominated by angular momentum zero states, thus we are dealing with a $^3P_0
\to$ $^1S_0 s$ transition.  Consequently, one deduces from unitarity
\beq
{\cal A} = |{\cal A}| \, e^{i\delta ({}^3P_0 )}~,
\eeq
with the $^3P_0$ $pp$ phase shift to be taken at the threshold energy in the
lab frame, $T^{\rm lab}_{\rm th} = m_\pi(2+m_\pi/2M) =279.65$~MeV, where
$\delta(^3P_0) = -6.3^\circ$ (FA95 solution of VPI). Thus the imaginary part
Im\,${\cal A}$ is  about $-1/9$ of the real part Re\,${\cal A}$ and contributes
negligibly to the total cross section near threshold proportional to $|{\cal
A}|^2$. The threshold T--matrix is a pseudoscalar, it is symmetric under the
exchange of the two ingoing protons $ \vec \sigma_1 \leftrightarrow \vec
\sigma_2,\, \vec p \to - \vec p$. Furthermore Eq.(\ref{T}) incorporates the
Pauli exclusion principle for the (indistinguishable) outgoing protons, since
left  multiplication with the spin exchange operator $(1+ \vec \sigma_1  \cdot
\vec \sigma_2)/2$ leads to a minus sign by the identity: ${1\over 2} (1+\vec
\sigma_1 \cdot \vec \sigma_2)\, ( i\, \vec \sigma_1 -i\, \vec \sigma_2 ,
\vec \sigma_1 \times \vec \sigma_2)  = - (\vec \sigma_1 \times \vec \sigma_2,
i\, \vec\sigma_1 -i\,\vec \sigma_2)$. Diagrams with crossed proton lines are
therefore automatically included. Approximating the near threshold T--matrix by
the T--matrix exactly at threshold one gets for the unpolarized total cross
section
\beq\label{stots}
\sigma_{\rm tot}(T_{\rm lab}) = \frac{M^4 \mu \sqrt{4+\mu}}{16 \pi^2
(2+\mu)^{9/2}}\, |{\cal A}|^2 \, (T_{\rm lab} - T_{\rm lab}^{\rm th}
)^2~,
\eeq
with $\mu = m_\pi / M$. Note that the flux and three-body phase space factors
have been approximated in Eq.(\ref{stots}) by an analytical expression which is
accurate within a few percent in the threshold region. Such a form has already
been proven to be quite accurate in chiral perturbation theory studies of the
reactions $\gamma p \to \pi^0 \pi^0 p$~\cite{bkm2pi0} and $\pi N \to \pi \pi
N$ \cite{bkmppn2},  where $\gamma$  denotes a real photon.
In the case of $pp\to pp\pi^0$, however, such an approximation is not
sufficient near threshold. This can be seen e.g. by taking the near threshold
data of refs.~\cite{meyer,bondar} and  dividing them by the three-body phase
space factor $\sim (T_{\rm lab}- T_{\rm lab}^{\rm th})^2$, i.e. $\sigma_{\rm
tot}(T_{\rm lab}) = C \cdot (T_{\rm lab} - T_{\rm lab}^{\rm th} )^2$. The
resulting values for $C$ are not constant in energy. This, of course, has to do
with the strong $pp$ final--state interaction in the $^1S_0$ partial wave. So
before we can extract a value for $|{\cal A}|$, we have to correct for the
final--state  interaction.

\subsection{Treatment of the final--state interaction}

The final--state interaction in the $^1S_0$ di--proton state modifies the
simple phase--space formula Eq.(\ref{stots}). We follow here a procedure
derived by  Watson~\cite{watson} where one essentially assumes that
final--state interaction is taking place only when the nucleons are on their
mass--shell. In this approach, the unpolarized total cross section for $ pp\to
pp \pi^0$  including final--state interaction takes the form
\begin{eqnarray}\label{stotm}
\sigma_{\rm tot}(T_{\rm lab}) &=& |{\cal A}|^2
\, \Big({M \over
4\pi}\Big)^3  {2 \sqrt{T_{\rm lab}} \over (2M+T_{\rm lab})^{3/2}} \nonumber \\
& &\times   \int_{2M}^{W_{\rm max}} d W  \sqrt{(W^2-4M^2)\,
\lambda(W^2,m_\pi^2,4M^2+2M  T_{\rm lab} )}\, F_p(W)~.
\end{eqnarray}
Here, $F_p(W)$ is the correction factor due the final--state interaction. We
evaluate it in the  effective range approximation. This is of course a very
strong assumption but it allows to explain the energy dependence of the
experimental total cross sections very accurately in terms of a single constant
amplitude ${\cal A}$. Separating off the final--state interaction
in that way, we can then pursue a diagrammatic approach to the (on-shell)
production amplitude ${\cal A}$. This will allow us to investigate in a simple
fashion the role of one-pion exchange and chiral loop effects together with
shorter range exchanges due to heavier mesons. For a more complete dynamical
description including off-shell final--state interactions and so on, a
microscopic
model treating also (unobservable) half--off shell effects is needed. After
these remarks, we return to the effective range approximation, in which the
function $F_p(W)$ takes the form
\begin{equation}\label{FW}
F_p(W) = {4 \sin^2 \delta_0(W) \over a_p^2 (W^2 -4M^2)} = \bigg\{1 +
{a_p\over 4} (a_p+r_p) (W^2-4M^2) + {a_p^2 r_p^2\over 64} (W^2-4M^2)^2
\bigg\}^{-1} \end{equation}
with $W$ the final--state di--proton invariant mass and
$\lambda(x,y,z)=x^2+y^2+ z^2 -2yz -2xz-2xy$ the K\"allen function. $W_{\rm max}
 = \sqrt{4M^2+2MT_{\rm lab} } -m_\pi $ is the kinematical endpoint of the
di--proton invariant mass spectrum. Furthermore, $a_p=(7.8098 \pm 0.0023)$~fm
and $r_p=(2.767\pm 0.010)$~fm, taken from  ref.\cite{nagels}, are the
scattering length and effective range parameter for elastic $pp$-scattering
including electromagnetic effects. Note that  we have fixed the
normalization of the correction factor $F_p(W)$ such that in the limit of
vanishing scattering length $a_p$ (i.e. vanishing final--state interaction),
$F_p(W)$ becomes identical to one (in the effective range
approximation). Furthermore, the condition $F_p(2M)=1$ ensures that there is
no final--state interaction effect exactly at threshold, as it must be
according to the definition of the threshold amplitude $\cal A$ in Eq.(1). In 
appendix C we give a simple derivation of $F_p(W)$ in scattering length
approximation (i.e. for $r_p=0$) using effective field theory methods.  

\begin{figure}[htb]
   \vspace{1.2cm}
   \epsfysize=7.9cm
   \centerline{\epsffile{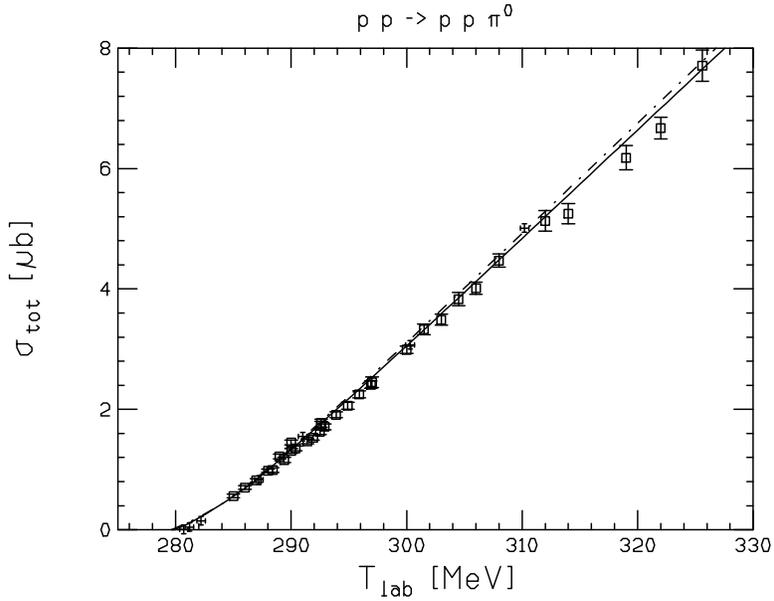}}
   \medskip
   \centerline{\parbox{13cm}{\caption{\label{figsig}
Fit to the total cross section for $pp \to pp\pi^0$ as described in the
text (solid line). The data are from {\protect{\cite{meyer}}} (boxes) and
{\protect{\cite{bondar}}} (crosses). The dashed line is explained in sec.~3.4.
}}}
\end{figure}

With the help of Eqs.(\ref{stotm},\ref{FW}) we are now in the position to
extract the values of $|{\cal A}|$ as shown in table~\ref{tab1}.
We ignore the data point at the lowest energy since that close to threshold
a better treatment of the infinite range Coulomb interaction in terms of proton
wave functions would be needed. The empirical value of ${\cal A}$ is thus
\beq\label{Aexp}
{\cal A}^{(\rm exp)} = (2.7- i\, 0.3)~{\rm fm}^4~,
\eeq
anticipating the positive sign of the real part from the calculation
in section~3.  Since we are mostly interested in achieving a qualitative
picture of the underlying production process, we refrain from assigning an
uncertainty to this number.  This number should be comparable to the one found
by Adam et al.~\cite{adam}, who employ  essentially the same method to correct
for the final--state interaction but use a different (unspecified)
normalization of the T--matrix. In Fig.\ref{figsig} we show the resulting total
cross  sections in comparison to the data of refs.\cite{meyer,bondar} using
Eqs.(\ref{stotm},\ref{FW},\ref{Aexp}).
\begin{table}[hbt]
\begin{center}
\begin{tabular}{|c|ccccccccc|}
    \hline\hline
    $T_{\rm lab}$~[MeV] & 282.2 & 285 & 290 & 295 & 300.3 & 308
                        & 314 & 319 & 325.6 \\
    \hline
    $\sigma_{\rm tot}$~[$\mu$b] & 0.148 & 0.56 & 1.31 & 2.06 & 3.07 & 4.47
                        & 5.25 & 6.18 & 7.71 \\
    \hline
    $|{\cal A}|$~[fm$^4$] & 2.42 & 2.73 & 2.69 & 2.64 & 2.70 & 2.71
                        & 2.64 & 2.66 & 2.73 \\
    \hline\hline
  \end{tabular}
\caption{Extracted values for the threshold amplitude $|{\cal A}|$ for
    the proton laboratory kinetic energies $T_{\rm lab}$ and cross sections
  $\sigma_{\rm tot}$  of ref.{\protect{\cite{meyer}}}.}\label{tab1}
\end{center}
\end{table}

\section{Meson--exchange contributions}

In this section, we work out a variety of one--boson exchange contributions
related to chiral (pion) and non--chiral (vector meson) physics to the
amplitude ${\cal A}$.

\subsection{Tree level Goldstone boson contributions}

\begin{figure}[htb]
   \epsfysize=5.5cm
   \centerline{\epsffile{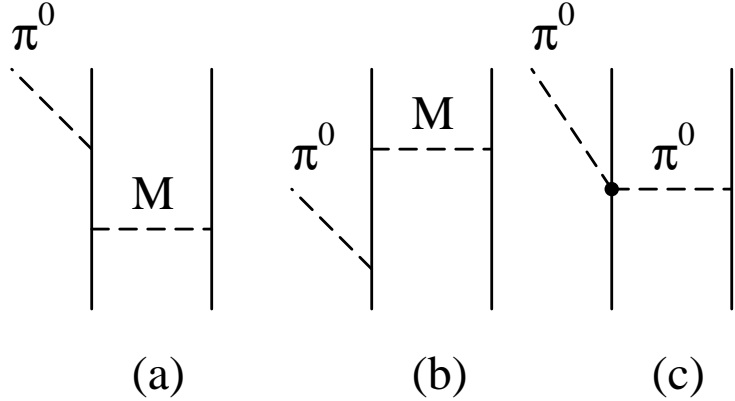}}
   \medskip
   \centerline{\parbox{13cm}{\caption{\label{fig2}
Feynman graphs for neutral pion production. (a) and (b) are the direct
terms, with meson exchange ${\cal M} = \pi^0, \eta, \omega, \rho^0$. (c) is the
rescattering graph. The heavy dot denotes the second order isoscalar chiral
$\pi\pi NN$--vertex. Graphs where the pion (dashed line)
is emitted from the other proton (solid) line and graphs with crossed
outgoing proton lines are not shown.   }}}
\end{figure}

In this and the following paragraph, we consider tree level and one--loop pion
exchanges contributing to the $pp\pi^0$ production process at threshold.
Due to the large proton momentum even at threshold, $|\vec p\,| \simeq
\sqrt{Mm_\pi}$, it is not appropriate to employ the frequently used heavy
baryon formalism for the nucleons. Let us demonstrate the failure of the heavy
baryon approach for a simple (but generic) example. Consider diagram
b) in Fig.\ref{fig2}
which involves a propagating nucleon after emission of the real $\pi^0$. We
show that this nucleon propagator can not be expanded in powers of $1/M$ in the
usual way. Let $v^\mu=(1,\vec{0} \,)$ be the four--vector which selects the
center--of--mass frame. The four--vector of the propagating nucleon is
$M  v + k$ with $k^\mu=
(-m_\pi/2, \vec p\,)$  and $k^2 = - Mm_\pi$. We start on the left hand side
with the correct relativistic result and then perform the usual $1/M$
expansion of the heavy baryon formalism,
\begin{equation}
-{1\over m_\pi} = {2 M\over (M  v +k)^2 - M^2} = {1\over
v\cdot k +k^2/2M} = {1\over v\cdot k} \sum_{n=0}^\infty \bigg( {- k^2 \over 2M
\, v\cdot k} \bigg)^n = - {2 \over m_\pi} \sum_{n=0}^\infty (-1)^n~.
\end{equation}
On sees that infinitely many terms of the  $1/M$--expansion contribute to
the same order. The resulting series does not even converge and oscillates
between zero and twice the correct answer. The source of this problem is the
extreme kinematics of the reaction $NN\to NN \pi$ with $|\vec p\,| \simeq
\sqrt{Mm_\pi}$. In that case the leading order operator ${\cal O}^{(1)} = i\,
v \cdot \partial $ and the next--to--leading order operator ${\cal O}^{(2)} =
- \partial \cdot \partial/2M$ lead to the same result, here, $\pm m_\pi/2$.
The heavy baryon formalism (and therefore also HBChPT) can not cope with
external momenta as large as $|\vec p\,| \simeq \sqrt{ Mm_\pi}$. This problem
is merely related to ``trivial'' kinematics and can be immediately overcome in
fully relativistic calculations. In the latter one does not presume that
vertices and propagators can be consistently expanded in inverse powers of the
large  nucleon mass $M$. The occurrence of such non--expandable nucleon
propagators is generic to $NN\to NN \pi$ and they will of course also enter in
loop diagrams. In order to treat correctly the kinematics, one has to evaluate
the loops relativistically, even though this may spoil the one--to--one
correspondence between the loop and small momentum/quark--mass expansion. A
consistent power counting scheme for processes $NN \to NN \pi$ involving loops
of arbitrary high order is not in sight at the moment.

\medskip

We thus turn to the relativistic description of the chiral pion--nucleon
system. Instead of giving the Lagrangian (see e.g. ref.\cite{gss}), we mention
only the relevant pion--nucleon vertices. The $\pi NN$--vertex (and also the
$\eta NN$-vertex) is of pseudovector type as required by chiral symmetry. The
second order isoscalar chiral $\pi\pi NN$ contact interaction, $\pi^0(q_1)
+N(p_1) \to \pi^0(q_2)+ N(p_2) $ reads,
\beq
{i \over f_\pi^2 } \bigg\{ -4 c_1 m_\pi^2 + 2c_3 q_1\cdot q_2 + {c_2'\over
2M} \Big[ (p_1+p_2)\cdot q_1 \, \barr q_2 +(p_1+p_2)\cdot q_2 \, \barr q_1\Big]
+ {c_2'' \over 2M^2} (p_1+p_2)\cdot q_1\, (p_1+p_2)\cdot q_2
\bigg\}\,.
\eeq
This form is unique since on mass--shell it gives the most general second
order $(s\leftrightarrow u)$ crossing symmetric polynomial contribution to the
invariant isoscalar $\pi N$ amplitudes,
\begin{equation} A^+(s,u) = {1\over f_\pi^2} \Big\{ -4c_1 m_\pi^2
+c_3( s+u-2M^2)+{c_2''\over 8M^2} (s-u)^2 \Big\}\,, \quad B^+(s,u) =
{c_2' \over 2 Mf_\pi^2} (s-u)\,, \end{equation}
with $s$ and $u$ the usual Mandelstam variables. The low--energy constants
$c_1, \,c_2',\,c_2'',\,c_3 $ have already been determined (at tree level)
in~\cite{bkmppn1} from low--energy $\pi N$ data and we list their values for
completeness:
\beq
c_1=-0.64\, , \quad c_2'=-5.63 \, , \quad
c_2''=7.41\,, \quad  c_3 = -3.90 \,\, ,
\eeq
all given  in GeV$^{-1}$. As it was shown in ref.~\cite{bkmlec}, the numerical
values of most of these low--energy constants  ($c_2', c_2'', c_3$) can be
understood largely from intermediate  $\Delta$ excitations.

\medskip

Consider first one--pion ($\pi^0$) exchange. The respective diagrams are shown
in Figs.~\ref{fig2}a,b (with ${\cal M} = \pi^0$).
We stress that these are relativistic Feynman
graphs, i.e. in the intermediate states they contain the full relativistic
fermion propagator which sums up several time-orderings. We also note that the
graphs  with the pion exchange after the emission of the $\pi^0$ from one of
the proton lines does not belong to the final state--interaction according to
our treatment (as an on--shell NN--process).  We find
\begin{equation} \label{pidir}
{\cal A}^{(\pi, {\rm dir})} = {g_{\pi N}^3 \over 4 M^4 (1+\mu) (2+\mu)} = 0.48
\, {\rm fm}^4~,
\end{equation}
with $g_{\pi N} = 13.4$ the strong pion--nucleon coupling constant.
We remark that it is often claimed that the $pp \pi^0$ final--state should be
suppressed due to chiral symmetry. This argument is based on the assumption of
the exchanged pion being soft, which, however, is not the case. The $\pi^0 p$
amplitude with one $\pi^0$ off its mass--shell  is only of linear order in the
pion mass $m_\pi$ (since $k^2=- Mm_\pi$) and this factor of
$m_\pi$ is cancelled by the pion propagator $[m_\pi(M+m_\pi)]^{-1}$.
Consequently, as Eq.(\ref{pidir}) shows, the threshold amplitude ${\cal A}$ in
$1\pi$--exchange approximation does not vanish in the chiral limit as it is
often claimed. In fact, chiral symmetry does not distinguish the $pp \to
pp\pi^0$ process from the other $NN \to NN\pi$  channels. Next, there is the
so--called  $\pi^0$ rescattering, as shown in Fig.~\ref{fig2}c,
\begin{equation} \label{pire}
{\cal A}^{(\pi, {\rm res})} = {g_{\pi N}\,\mu  \over f_\pi^2 M (1+\mu)}
\Big[ {c_3\over 2} +\Big(1+{\mu \over 4}\Big) c_2'+\Big(1+{\mu \over 4 }
\Big)^2 c_2''-2 c_1  \Big] = 0.46\, {\rm fm}^4~,
\end{equation}
with $f_\pi = 92.4\,$MeV the pion decay constant. Again, there is a marked
difference to the heavy baryon case. To leading order, the relativistic
couplings $c_2'\,, \, c_2''$ combine to give the $c_2=c_2'+c_2''$ term in the
heavy baryon approach. In previous HBChPT calculations the $c_2$ term was
found with an incorrect prefactor $1/2$. The relative factor of two in the
relativistic calculation comes from the fact that products of nucleon and pion
four-momenta are not dominated anymore by the term nucleon mass times pion
energy. For the reaction $NN \to NN \pi$ the product of the nucleon and pion
three-momenta can be equally large, since $|\vec p\,| \simeq \sqrt{Mm_\pi}$.

Interestingly, the combination of low--energy constants in Eq.(\ref{pire}) is
dominated (to about $90\%$) by the last term $\sim -2c_1$, which is related to
the
so--called pion--nucleon sigma--term, $\sigma_{\pi N}(0)= -4 c_1 m_\pi^2$ (to
leading order). Therefore the strength of the $\pi^0$--rescattering (at
threshold) is almost entirely due to this particular chiral symmetry breaking
term. The effects from the $\Delta(1232)$--resonance encoded in the low--energy
constants $c_2',\,c_2'',\,c_3$ turn out to be very small. In order to check
this interpretation, we have evaluated the contributions from explicit
$\Delta(1232)$--excitations, using the Rarita--Schwinger formalism and the
well--satisfied coupling constant relation $g_{\pi N \Delta} = 3 g_{\pi
N}/\sqrt2$. We find:
\begin{eqnarray}\label{Adel}
{\cal A}^{(\Delta)} &=& {g_{\pi N}^3 \, \mu \over 16
M_\Delta^2 M^3 (1+\mu)} \bigg\{ { M^2(4+\mu)(M_\Delta+M(1-\mu))\over M_\Delta^2
+M^2(2\mu-1) } \nonumber \\ & & +(2Z-1) \Big[ 4(1+Z)M_\Delta -M
(2Z(3+\mu)+1+\mu) \Big] \bigg\} = 0.04\, {\rm fm}^4~,
\end{eqnarray}
where we used for the off--shell parameter $Z=-0.3$, the value which maximizes
the $\Delta(1232)$--contribution to the $P_{33}$ $\pi N$ scattering volume.
Such a
value of $Z$ is also consistent with neutral pion photoproduction off protons.
The small value of ${\cal A}^{(\Delta)}$ confirms the interpretation of the
rescattering contribution Eq.(\ref{pire}) given above.

\medskip

The next Goldstone boson which can contribute is the $\eta (547)$. Consider the
graphs in Figs.~\ref{fig2}a,b with ${\cal M} = \eta$. We find
\begin{equation}\label{Aeta}
{\cal A}^{(\eta)} = {g_{\pi N} g_{\eta N}^2 \, \mu \over 4M^2 (m_\eta^2
+M^2 \mu) (2+\mu) }= 0.02 \, {\rm fm}^4~,
\end{equation}
where we have employed the SU(3) value for the $\eta N$ coupling constant
together with the simplified ratio of the octet axial vector coupling constants
$D/F=1.5$, which leads to $g_{\eta N} = 4.6$. Since this contribution is tiny,
the precise value of this coupling does not matter.

\subsection{Pion loop effects}

We do not attempt a full one--loop calculation here, but rather consider
certain (simple) classes of loop graphs which are genuine to the process under
consideration. We use dimensional regularization and minimal subtraction to
eliminate divergences and set the renormalization scale equal to the proton
mass $M$. For estimating the genuine size of pion loop effect, such a procedure
ignoring renormalization via counterterms should be sufficient and the
resulting numbers should be considered indicative.

\begin{figure}[htb]
   \vspace{0.9cm}
   \epsfxsize=12cm
   \centerline{\epsffile{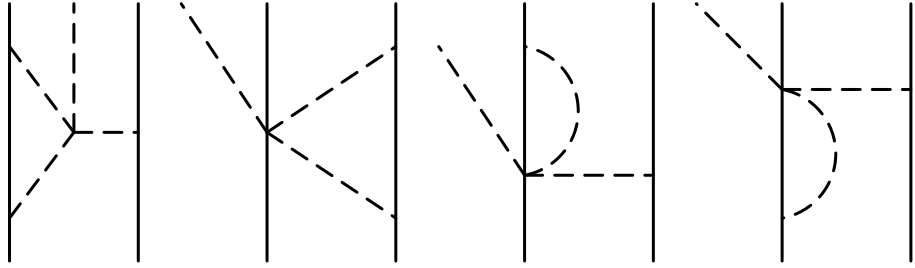}}
   \medskip
   \centerline{\parbox{13cm}{\caption{\label{fl1}
Class of one loop graphs involving the $\pi\pi$ interaction.
For further notation, see Fig.3. }}}
\end{figure}

\begin{figure}[htb]
   \epsfysize=7cm
   \centerline{\epsffile{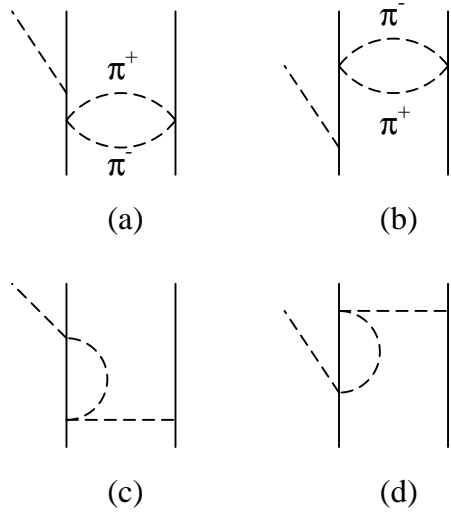}}
   \medskip
   \centerline{\parbox{13cm}{\caption{\label{fl2}
Further loop diagrams proportional to $g_{\pi N}/f_\pi^4$. (a) and (b)
represent $\pi^+\pi^-$ exchange between protons, (c) and (d) are the $\pi^+$
rescattering diagrams. For further notation, see Fig.3.  }}}
\end{figure}

\begin{figure}[htb]
   \epsfxsize=10cm
   \centerline{\epsffile{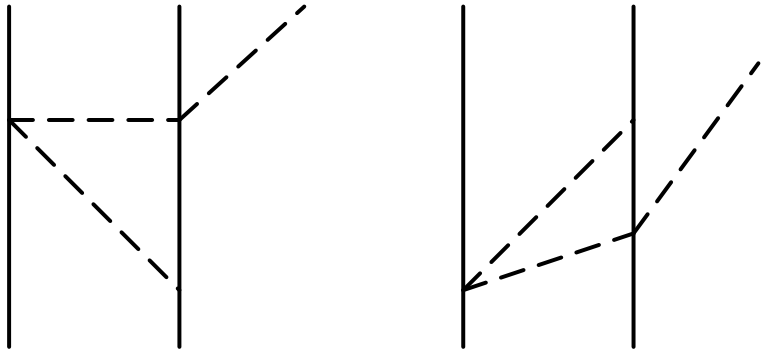}}
   \medskip
   \centerline{\parbox{13cm}{\caption{\label{fl3}
Loop diagrams with $\pi^0$ emission via charge exchange. For further notation,
see Fig.3. }}}

\end{figure}

Consider first a certain class of pion loop diagrams which involve the
$\pi\pi$ interaction, compare Fig.~\ref{fl1}. Notice that  only the full
class of diagrams is independent of the choice of the interpolating pion field.
Within the calculational scheme mentioned above, this class of graphs
gives
\begin{eqnarray}\label{loop1}
{\cal A}^{{(\rm loop,1})}
&=& {g_{\pi N}^3 (2+\mu)(1-\mu) \over (8\pi M
f_\pi)^2 (1+\mu)} \bigg\{ \ln \mu -{1\over2}+\sqrt{1+4\mu}\,\ln{1+\sqrt{1+4\mu}
\over 2\sqrt{\mu}} \nonumber \\ & & +\int_0^1dx \int_0^x dy {y\over
y^2+\mu^2(1-y) +\mu (1-x)(x-y)} \bigg\} = -0.10 \, {\rm fm}^4~.
\end{eqnarray}
We remark that the expression in the curly brackets is not singular
in the chiral limit $\mu \to 0$. It has the following $\mu$-expansion:
$ -1/2+ \sqrt{\mu}\, \pi^2/8 +{\cal O}(\mu \ln \mu)$.
Again, one faces here the usual problem of relativistic loops. ${\cal A}^{({\rm
loop,1})}$ does not vanish in the chiral limit $\mu\to 0$  and is therefore not
suppressed by powers of the pion mass $m_\pi$ compared to tree
graphs. Nevertheless, its numerical value is small. The threshold amplitude
${\cal A}^{({\rm loop,1})}$ is actually proportional to the relativistic
loop contribution~\cite{gss} to the nucleon scalar form factor evaluated at an
invariant  momentum transfer squared $t=-Mm_\pi$,
\beq
{\cal A}^{(\rm loop,1)} = {g_{\pi N} (2+\mu)(\mu-1) \over 12 M f_\pi^2
m_\pi^2 (1+\mu) } \,\sigma_{\pi N}(-M^2\mu)_{\rm loop}~.
\eeq
Whereas ref.\cite{isra} claimed that these loop graphs were sizeable
and are essential for a quantitative description of the data, we find
that they give only a small $-4\%$ correction compared to the empirical value
of ${\cal A}$. The reason for this discrepancy is to be found in the
inappropriate application of the heavy baryon formalism to $pp\to pp\pi^0$ in
ref.\cite{isra}. Next, we consider the entire class of loop graphs proportional
to $g_{\pi N}/f_\pi^4$, see graphs in Fig.\ref{fl2} and Fig.\ref{fl3}.
Figs.\ref{fl2} (a) and (b) represent   $\pi^+\pi^-$ exchange  between protons
(in form of a bubble diagram)  before or after the emission of the neutral
pion. We find:
\beqa \label{loop2}
{\cal A}^{(\rm loop,2)}
&=& {g_{\pi N}(2-\mu)\mu \over 768\pi^2
f_\pi^4 (2+\mu)} \bigg\{(1+6\mu) \ln \mu - {5\over 6} -4\mu +(1+4\mu)^{3/2}\,
\ln{1+\sqrt{1+4\mu} \over 2\sqrt{\mu}} \bigg\}\nonumber \\
&=& -0.01 \, {\rm fm}^4~.
\eeqa
Note that these loop corrections vanish in the chiral limit $\mu \to 0$. In
the class of diagrams proportional to $g_{\pi N}/f_\pi^4$  there is also
$\pi^0$--exchange between the protons with  $\pi^+$ rescattering on one of the
nucleons, see (c) and (d) in Fig.\ref{fl2}. These graphs give
\begin{eqnarray} \label{loop3}
{\cal A}^{(\rm loop,3)} &=& { g_{\pi N}\, \mu \over [8\pi
f_\pi^2(1+\mu) (1-2\mu)]^2} \bigg\{(1+\mu)(1-2\mu)(4-6\mu-\mu^2)+(11-16
\mu+\mu^2+\mu^3) \nonumber \\ & & \cdot  \mu^2 \ln \mu + \mu (1+\mu) (2-2\mu-
\mu^2) \sqrt{\mu (4+\mu)} \,\ln{2+\mu+\sqrt{\mu(4+\mu) } \over 2}
\bigg\} \nonumber \\ &=&   0.22\, {\rm fm}^4~\,\,.
\end{eqnarray}
This contribution vanishes again in the chiral limit $\mu \to 0$.
Finally, there are graphs where a $\pi^+\pi^-$--pair is emitted from one
proton  and the $\pi^0$ emission proceeds via charge exchange from the
other nucleon line, see Fig.\ref{fl3}. Straightforward evaluation of these
diagrams gives
\begin{eqnarray}\label{loop4}
{\cal A}^{(\rm loop,4)} & & = {g_{\pi N}\,\mu \over (8\pi f_\pi^2)^2 } \bigg\{
{\mu(9\mu-2-\mu^2)\over 2(1+\mu) (1-2\mu)}
+{6-12 \mu +19\mu^2 +2\mu^3 +\mu^4  \over 2(1+\mu)^2 (1-2\mu)^2}
\mu^2 \ln \mu \nonumber \\ & & + {\mu^2 (2-\mu)
\over 2(1-2\mu)^2} \sqrt{\mu (4+\mu)} \ln{2+\mu +\sqrt{\mu(4+\mu)} \over 2} +
\int_0^1 dx \int_0^x dy \, y \times \nonumber \\ & &  \bigg[ {2x-2-5y
+\mu(3-3x-
y) + \mu^2(3-x) \over y^2 +\mu (1-x)(x-3y) +\mu^2(1-2y+xy) }
+{2-2x-3y+\mu(x-1)+\mu^2 \over y^2 +\mu(1-x)(x+y) +\mu^2(1-y) } \bigg] \bigg\}
\nonumber \\ &&= -0.25 \, {\rm fm}^4~\,\,,
\end{eqnarray}
a contribution which again vanishes in the chiral limit. Summing all these
various loop contributions in
Eqs.(\ref{loop1},\ref{loop2},\ref{loop3},\ref{loop4}), we find
\beq\label{Aloop}
{\cal A}^{(\rm loop)} =  -0.14 \, {\rm fm}^4~,
\eeq
which is a rather small number as compared to the tree level pion exchange
or the vector meson contributions (section 3.3). We observe sizeable
cancelations between the various classes of loop graphs and conclude therefore
that these do not  play a significant role for explaining the threshold
amplitude ${\cal A}$. The imaginary part of ${\cal A}$ is discussed in
appendix B.

\subsection{Heavy meson exchanges}

Due to the large momenta involved, one expects additional contributions
due to the exchange of heavier mesons, which are of much shorter range
than the pion exchange considered so far. From the symmetry point of
view such terms are much less constrained and thus exhibit a certain
unavoidable model--dependence. Nevertheless, these terms can play
a significant role as first stressed by Lee and Riska~\cite{LR} and confirmed
by  Horowitz et al.~\cite{chuck}. We do not discuss here additional effects
from  vector meson nucleon form factors, which are a model--dependent (and
unobservable) concept to account for the finite size of the hadrons involved.
Note also that in a strict field--theoretical sense such form factors can not
be uniquely defined. At the invariant momentum transfer $t =
-0.127$ GeV$^2$ the form factor effect is not expected to be large.
For example, one expects a $10\%$ effect for typical
monopole form factors with cut--offs $\Lambda_{\omega,\rho} \simeq 1.5\,$GeV.

\medskip

Consider first neutral vector mesons. We start with the
$\omega(782)$. There is sizeable uncertainty about its coupling constant
to the nucleon, extreme values are e.g. found in the dispersion--theoretical
analysis of  the nucleons electromagnetic form factors, $g_{\omega N} \simeq
21$~\cite{MMD}. However, it is not clear how one has to transcribe such
a value to the one--boson exchange picture of the NN force. In conventional
boson--exchange models, the inclusion of the correlated $\pi\rho$ continuum
allows one to work with a coupling constant that is compatible with the
SU(3) value,  $g_{\omega N} \simeq 9$ or the value $g_{\omega N} = 10.1 \pm
0.9$ found from forward NN--dispersion relations \cite{kroll}. For a detailed
discussion,  see e.g. ref.~\cite{hol}. There is agreement that the
tensor--to--vector coupling ratio of the $\omega$-meson is very small. If we
set
$\kappa_\omega =0$ and use the coupling constant $g_{\omega N}=10$, we get
\begin{equation} \label{Aom}
{\cal A}^{(\omega)} = {g_{\pi N}\, g_{\omega N}^2 (2-\mu) \over 2 M^2
(M_\omega^2 +M^2 \mu) (2+\mu) } = 1.35 \, {\rm fm}^4~,
\end{equation}
which is quite sizeable. In a similar fashion, we evaluate the $\rho (770)$
contribution. Here, there is less debate about the coupling constant $g_{\rho
N}$ and also, it is well established that the tensor--to--vector coupling
ratio $\kappa_\rho$ is large. For simplicity, we use $g_{\rho N} =3$ (obtained
from $\rho$-universality, $g_{\rho N} = g_{\rho}/2$ with $g_{\rho}=6$) and
$\kappa_\rho =6$. That leads to
\begin{equation} \label{Arho}
{\cal A}^{(\rho)} = {g_{\pi N}\, g_{\rho N}^2 \over 2 M^2
(M_\rho^2 +M^2 \mu) (2+\mu) } \Big[ 2+\mu (\kappa_\rho^2-\kappa_\rho-1)
+\mu^2  \kappa_\rho \Big( 1+{9\over 8} \kappa_\rho\Big) \Big] = 0.48 \, {\rm
fm}^4~.
\end{equation}
We note that this form is very different from what has been used in the
literature so far, where one finds the $\omega,\rho$--exchange to be
proportional to $\mu (1+\kappa_{\omega,\rho})$. We do not employ any
inappropriate non--relativistic approximation here and thus obtain the results
shown in Eqs.(\ref{Aom},\ref{Arho}). Note that due to the large
tensor--to--vector coupling ratio of the $\rho$-meson $\kappa_\rho=6$ the terms
proportional to $\mu$ and $\mu^2$ in the square bracket of Eq.(\ref{Arho}) are
most important. We have also investigated the role of $\phi (1020)$ exchange,
which can be related to the strangeness content of the nucleon wave function.
The values of the $\phi N$ coupling constant span a large range, as
documented in table~2 of ref.\cite{mmsv}. To get a more precise number, we
proceed as follows. We assume that the tensor--to--vector coupling ratios
are given by the dispersive analysis of the nucleon electromagnetic
form factors, $\kappa_\rho = 6.1$, $\kappa_\omega = -0.16$ and $\kappa_\phi
= -0.22$~\cite{MMD}. Using furthermore the SU(3) relation $g_{\omega N} =
3g_{\rho N} - \sqrt{2}g_{\phi N}$~\cite{hol} together with $g_{\rho N} =
2.63$~\cite{hopi} and $g_{\omega N} = 10.1$~\cite{kroll}, we have
$g_{\phi N} = -1.56$. This leads to ${\cal A}^{(\rm V)} = {\cal A}^{(\rho)} +
{\cal A}^{(\omega)} + {\cal A}^{(\phi)} = (0.38 + 1.40 + 0.02)$ fm$^4 = 1.80$
fm$^4$, which is practically identical to the result obtained above using the
simplified coupling constants. Thus, the sum of the vector meson contributions
is fairly stable against parameter variations and also the $\phi(1020)$ does
not play any role. Finally, we remark that substituting $(g_{\rho N}, M_\rho,
\kappa_\rho)$ by $(e, 0, \kappa_p)$ one can convince oneself that the
one--photon exchange  gives a tiny correction of ${\cal A}^{(\gamma)} = 0.0085$
fm$^4$, as it is of course expected.

\medskip

Another mechanism, first proposed in this context in~\cite{bira2}, is the
emission of the neutral pion from the anomalous $\omega \rho\pi$-vertex, with
the $\rho(770)$ coupling to one and the $\omega(782)$ to the other proton.
This type of graph is similar to well-known meson--exchange currents in
electromagnetic processes. The pertinent interaction vertex, with its
strength given by the coupling constant $G_{\omega\rho\pi}$,
can be determined from the  anomalous Wess--Zumino--Witten term for vector
mesons (we only show the part  of relevance here),
\beq \label{WZW} {\cal L}_{\omega \rho \pi} = -{G_{\omega \rho
\pi}\over f_\pi} \, \epsilon^{\mu \nu \alpha \beta} \,( \partial_\mu \,
\omega_\nu )\, \vec \rho_\alpha \cdot \partial_\beta \vec \pi~,
\eeq
with $\epsilon^{\mu \nu \alpha \beta}$ the totally antisymmetric tensor in
four dimensions ($\epsilon^{0123}=-1$). Using again the universal
$\rho$-coupling,
$g_\rho=6$, the gauged Wess--Zumino--Witten term for vector mesons
leads to the coupling constant
\beq
G_{\omega \rho \pi} = {3g_\rho^2\over 8\pi^2} = 1.37~.
\eeq
Similar values for $G_{\omega \rho \pi}$ have been found in
refs.\cite{jain,klingl} from systematic studies of $\omega(782)$-- and
$\phi(1020)$--decays. The contribution of the anomalous
$\omega \rho\pi$--vertex to ${\cal A}$ is given by
\begin{equation} \label{AWZW}
{\cal A}^{(\omega\rho\pi)} = {g_{\omega N} g_{\rho N}(1+\kappa_\rho)
G_{\omega\rho\pi} M \mu^2 \over 2 f_\pi (M_\omega^2+M^2 \mu)(M_\rho^2+M^2 \mu)
} = 0.09 \, {\rm fm}^4~,
\end{equation}
which is at first glance quite small. However, due to the two derivatives
appearing in Eq.(\ref{WZW}), one expects this contribution to be of much
larger importance in the P--wave amplitudes.
An analogous short range mechanism is the
$\pi^0$ emission from the $a_0 \eta \pi$-vertex, proposed in
ref.\cite{bira2}. We have evaluated the respective contribution to
${\cal A}$ and found that it is negligibly small, ${\cal A}^{(a_0 \eta
  \pi)} = 0.003$ fm$^4$. Finally, we do not consider a
scalar meson exchange here. First, there is no scalar meson resonance which
couples strongly to the nucleon and secondly, the fictitious ``$\sigma(550)$''
of one--boson exchange models just simulates the long and intermediate range
part of uncorrelated $2\pi$--exchange in the NN interaction \cite{nnpap}. The
latter comes along with pion loops and is to some extent contained in the loop
graphs considered above.

\subsection{Total threshold amplitude}

We are now in the position to evaluate the full amplitude ${\cal A}$
from the various contributions. Combining
Eqs.(\ref{pidir},\ref{pire},\ref{Aeta},\ref{Aloop},\ref{Aom},\ref{Arho},\ref{AWZW}),
we get
\beqa
{\cal A}^{(\rm thy)} &=& {\cal A}^{(\pi, \rm dir)} + {\cal A}^{(\pi, \rm res)}
+ {\cal A}^{(\eta)} + {\cal A}^{(\rm loop)} + {\cal A}^{(\omega)} + {\cal A}^{(
\rho)} + {\cal A}^{(\omega\rho\pi)} \nonumber \\
&=& (0.48 \,\,\,\, + 0.46 \,\,\,\,+ 0.02
\,\,\,\,- 0.14\,\,\,\, + 1.35\,\, + 0.48 + 0.09)~{\rm fm}^4\nonumber\\
&=& 2.74~{\rm fm}^4~,
\eeqa
which compares well with the empirical value given in Eq.(\ref{Aexp}).
The resulting total cross section $\sigma_{\rm tot}(T_{\rm lab})$ is shown in
Fig.~\ref{figsig} by the dashed--dotted line. Of course, taken the uncertainty
in certain coupling constants and our simplified treatment of the final--state
interaction,  the 1\% agreement between our theoretical prediction
for ${\cal A}$ and its empirical value, Eq.(\ref{Aexp}), should not be
taken too serious. We only want to make the point that these
well--known boson--exchange diagrams, when evaluated fully
relativistically, can explain the near threshold data for
$pp \to pp \pi^0$. Note also that  the contribution from explicit
$\Delta(1232)$-excitation, cf. Eq.(\ref{Adel}), is contained in the
$\pi^0$ rescattering
term via the low--energy constants $c_2',\, c_2'',\, c_3$.

\section{Charged pion production in $pp$--collisions}

In this section we will discuss charged pion production using the same
approach. The T--matrix for charged pion production in proton--proton
collisions, $p_1(\vec p\,) +p_2(-\vec p\,) \to p+n +\pi^+$, at threshold in the
center--of--mass frame reads (see also appendix A for the general $NN\to
NN\pi$ threshold T--matrix),
\begin{equation} \label{thpiplus}
{\rm T}^{\rm cm}_{\rm th}(pp\to pn \pi^+) = {{\cal A}\over \sqrt 2} \, (i\,\vec
\sigma_1 -  i\,\vec \sigma_2 + \vec \sigma_1 \times \vec \sigma_2)\cdot\vec p
- \sqrt 2 {\cal B} \, i(\vec \sigma_1+\vec \sigma_2) \cdot \vec p \,\,.
\end{equation}
As indicated by the ordering of the three particles $pn\pi^+$ in the final
state, the spin--operator $\vec \sigma_1$ is understood to be sandwiched
between the spin--states of the ingoing proton $p_1(\vec p\,)$ and those of
the outgoing proton, while $\vec \sigma_2$ acts between the proton $p_2(-\vec
p\,)$ and the outgoing neutron. Since the $pn$--system at rest can be in a
spin--singlet or in a spin--triplet state two different transitions are
possible at threshold. The threshold amplitude for the singlet transition
$^3P_0 \to$ $^1S_0s$ is by isospin symmetry proportional (with a factor
$1/\sqrt2$) to the threshold amplitude ${\cal A}$ for the reaction $pp \to pp
\pi^0$ introduced in Eq.(\ref{T}). The new threshold amplitude for the triplet
transition $^3P_1 \to$ $^3S_1s$ is called ${\cal B}$ and the factor $-\sqrt 2$
was taken out for convenience in Eq.(\ref{thpiplus}). In analogy to Eq.(3) one
deduces from unitarity
\beq \label{unib} {\cal B} = |{\cal B}| \, e^{i \delta(^3P_1)}\,\,, \eeq
with the $^3P_1$ $pp$  phase shift to be taken at $T_{\rm lab}^{\rm th}=292.3$
MeV, where $\delta(^3P_1)= -28.1^\circ$ (FA95 solution of VPI). Because of
this larger phase fixed by unitarity the imaginary part Im\,${\cal B}$ will
contribute non--negligibly to the total cross sections near threshold. Notice,
that the abovementioned unitarity relation Eq.(\ref{unib}) neglects the
(small) inelasticity due to the $pp \to d \pi^+$ channel which opens $4.8$ MeV
lower at $T_{\rm lab}=287.5$ MeV.

\subsection{Extraction of the threshold amplitudes}

Employing the same method as in section~2.2 to correct for the strong
S-wave $pn$ final--state interaction, the unpolarized total cross section
for $ pp\to pn \pi^+$ reads,
\begin{eqnarray} \label{piplus}\sigma_{\rm tot}(T_{\rm lab}) &=& \Big({M \over
4\pi}\Big)^3   {2 \sqrt{T_{\rm lab}} \over (2M+T_{\rm lab})^{3/2}}
\nonumber \\
&&\times \int_{M+M_n}^{W_{\rm max}}{d W\over W} \sqrt{\lambda(W^2,M^2,M_n^2)\,
\lambda(W^2,m_{\pi^+}^2,4M^2+2M  T_{\rm lab} )} \nonumber \\
&&\times \Big\{ |{\cal A}|^2\, F_s(W)+2|{\cal B}|^2\, F_t(W) \Big\}~.
\end{eqnarray}
The correction factors from the $pn$ singlet and triplet S--wave final state
interaction are given in the effective range approximation by
\begin{equation}
F_{s,t}(W) = \bigg\{1 +a_{s,t} (a_{s,t}+r_{s,t})P^2_* +
{1\over 4}a_{s,t}^2 r_{s,t}^2 P^4_* \bigg\}^{-1}~,
\end{equation}
with $W$ the final--state proton--neutron invariant mass and $W_{\rm max} =
\sqrt{4M^2+2MT_{\rm lab} } -m_{\pi^+}$. $M$ still denotes the proton mass and
$M_n=939.57$ MeV stands for the neutron mass. The quantity $P^2_*=\lambda(W^2,
M^2, M_n^2)/4W^2$ is the squared $pn$ center--of--mass momentum. The singlet
and triplet scattering lengths and effective range parameters for elastic
$np$-scattering are taken from~\cite{nagels}, their empirical values being
$a_s=(23.748 \pm 0.010)\,$fm, $a_t=(-5.424 \pm 0.004)\,$fm, $r_s=(2.75 \pm
0.05)\,$fm and $r_t=(1.759 \pm 0.005)\,$fm. We neglect here the coupling
between the $^3S_1$ and the $^3D_1$ $pn$--states, which should be very small at
the energies under consideration. Such effects go beyond the accuracy of the
effective range approximation and our treatment of the $pn$ final--state
interaction.

The data base of total cross sections for the process $pp\to pn \pi^+$ in
the 30 MeV region above threshold consists at present of five data points
measured at IUCF \cite{pipn}. We leave out the data point at the highest energy
$T_{\rm lab}=319.2$ MeV where the $\pi^+$ angular distributions are no more
isotropic and thus P-waves start to become important.  We also found it
important to ignore the data point at the lowest energy $T_{\rm lab} = 294.3$
MeV. Using Eq.(\ref{piplus}) for the total cross section and the value of $|{
\cal A}| = 2.72$ fm$^4$ as determined from the $pp\to pp \pi^0$ data, one finds
in a least square fit of the remaining three data points for the modulus of
the triplet threshold amplitude,
\begin{equation}
|{\cal B}| = 3.16 \, {\rm fm}^4\,\, ,
\end{equation}
with a very small total $\chi^2= 0.044$. For these values of $|{\cal A}|$ and
$|{\cal B}|$ the data point at $T_{\rm lab} =319.2$ MeV is underestimated by
$17\%$ in the S-wave approximation (see table 2). An unconstrained
fit of the same three data points gives $|{\cal A}|=3.00$ fm$^4$ and $|{\cal
B}|=3.15$ fm$^4$ with a marginally smaller total $\chi^2=0.042$. It is quite
remarkable that $|{\cal A}|$ is found to be in $10\%$ agreement with the value
obtained from fitting the many precise near threshold
$pp \to pp \pi^0$ data. Due to the very
strong $pn$ final--state interaction in the $^1S_0$ exit channel ($a_s= 23.75$
fm) the singlet transition contributes a factor 30 to 40 less to the total
cross section than the triplet transition. From our fit we get at the lowest
energy $T_{\rm lab}=294.3$ MeV a total cross section of $0.57 \, \mu$b,
compared to the experimental value of $(0.71 \pm 0.04)\, \mu$b given in
ref.\cite{pipn}. Only if one widens the error band of this data point by
the $\pm 15 \%$ absolute normalization uncertainty it becomes (at the lower
end) marginally consistent with the remaining data. If, however, the data point
at $T_{\rm lab}= 294.3$ MeV were included in an unconstrained fit, very
different values of the threshold amplitudes, $|{\cal A}|= 6.76$ fm$^4$ and
$|{\cal B}|= 2.91$ fm$^4$, would be found with a total $\chi^2=1.0$.  In
particular $|{\cal A}|$ would be a factor 2.5 larger than the one obtained
from the fit to the $pp \to pp \pi^0$ data. Of course, such a large deviation
from isospin symmetry is unacceptable.
\begin{table}[hbt]
\begin{center}
\begin{tabular}{|c|ccccccc|}
    \hline\hline
    $T_{\rm lab}$~[MeV]  & 294.3 & 295.1$^{\,*}$ & 298.0$^{\,*}$ & 299.3 & 
    306.3 & 314.1 & 319.2 \\    \hline
    $\sigma_{\rm tot}^{\rm exp}$~[$\mu$b]
                   & $0.71$ & $1.1^{\,*}$ & $3.84{\,*}$
                   &  $4.81$ &  $13.91$ & 
                          $25.5$ & $41.1$ \\
    $\delta\sigma_{\rm tot}^{\rm exp}$~[$\mu$b]
                   & $\pm 0.04$ & $\pm0.3^{\,*}$ & $\pm 0.37^{\,*}$
                   &  $\pm 0.24$ &  $\pm 0.65$ & 
                          $\pm 1.6$ & $\pm 1.7$ \\
    \hline
    $\sigma_{\rm tot}^{\rm fit}$~[$\mu$b]
          & 0.57 & 1.07 & 3.46 & 4.81 & 13.82 & 25.76 &  34.14 \\
    \hline\hline
  \end{tabular}
\caption{Total cross sections for $pp\to pn\pi^+$ as a function of
$T_{\rm lab}$. The data are taken from ref.{\protect{\cite{pipn}}} and  the 
preliminary data points from COSY-TOF \cite{brink} are marked by an asterik. 
The (constrained) fit  is  described in the text.}\label{tab2}  \end{center}
\end{table}

Anticipating the positive sign of the real part from the calculation in the
following section and using the information from the $^3P_1$ phase
shift, one gets the following experimental value of the triplet threshold
amplitude ${\cal B}$,
\begin{equation} \label{Bemp}
{\cal B}^{(\rm exp)} = (2.8 - i\, 1.5 )\, {\rm fm}^4\,\,.
\end{equation}
This number should be considered indicative since the systematic error
of the extraction method is not under control when only three data points are
fitted. Notice also that the imaginary part is fairly sizeable, quite
different from ${\cal A}$, the threshold amplitude for $pp \to pp\pi^0$.

\subsection{Diagrammatic approach}

Next, we turn to the evaluation of the relativistic Feynman diagrams
contributing to $pp\to pn \pi^+$ at threshold. In addition to the ones
considered for $pp \to pp \pi^0$, there is now the possibility for isovector
pion--rescattering. The chiral $\pi N$ Lagrangian contains such vertices at
leading order (the so--called Weinberg--Tomozawa vertex) and at
next--to--leading order (a vertex proportional to the low--energy constant
$c_4=2.25$ GeV$^{-1}$). In fact there are always two isovector rescattering
diagrams, one with a $\pi^0$ and another one with a $\pi^+$ being exchanged
between the nucleons. The isospin factors of both diagrams are equal with
opposite sign. However, before adding them one has to account for the fact
that the role of $p$ and $n$ is interchanged in both graphs. This is done by
multiplying those graphs where the final state neutron $n$ comes from the
initial state proton $p_1(\vec p\,)$ with the negative spin--exchange operator
$-(1+ \vec \sigma_1 \cdot \vec \sigma_2)/2$.  Altogether, one finds from
isovector pion--rescattering at leading and next--to--leading order,
\begin{equation} \label{isovec} {\cal B}^{(\pi,{\rm iv})} = {g_{\pi N} ( c_4
m_\pi -1)\over 2 M^2 f_\pi^2 (1+\mu)} = -0.82\, {\rm fm}^4 \,\,. \end{equation}
Of course, we neglect here the small isospin breaking due the different charged
and neutral pion masses and the different proton and neutron masses.
>From the other pseudoscalar meson ($\pi$ and $\eta$) exchange diagrams, one
finds the following contributions to the triplet amplitude,
\begin{eqnarray} \label{bpidir}
{\cal B}^{(\pi,{\rm dir})} &=& {g_{\pi N}^3 (3+2\mu) \over 4M^4(1+\mu)(2+\mu) }
 = 1.58\, {\rm fm}^4 \,\,, \\ \label{isoresc}
{\cal B}^{(\pi,{\rm res})} &=& {\cal A}^{(\pi,{\rm res})} = 0.46 \, {\rm fm}^4
\,\,, \\ \label{beta} {\cal B}^{(\eta)} &=& -{\cal A}^{(\eta)} = -0.02 \, {\rm
fm}^4 \,\,, \end{eqnarray}
(see also Eqs.(\ref{pire},\ref{Aeta})). Note that the (direct) $1\pi$--exchange
contribution ${\cal B}^{(\pi,{\rm dir})}$ is rather large due to an enhancement
factor $3+2\mu$ in comparison to ${\cal A}^{(\pi,{\rm dir})}$ given in
Eq.(\ref{pidir}). The dominance of the chiral one--pion exchange in the triplet
transition $^3P_1 \to$ $^3S_1 s$ found here supports the argument of
\cite{hhhms} concerning the dominant role of the long range (chiral) pion
exchange in the reaction $pp \to d \pi^+$ which proceeds via the same
transition near threshold.

>From the vector meson  ($\rho$ and $\omega$) exchange diagrams one finds the
following contributions  to ${\cal B}$,
\begin{eqnarray} \label{vmeson1}
{\cal B}^{(\omega)} &=& {g_{\pi N} g_{\omega N}^2 \over M^2 (M_\omega^2 +M^2
\mu) (2+\mu)} = 1.46\, {\rm fm}^4\,\,,  \\ \label{vmeson2}
{\cal B}^{(\rho)} &=&  {g_{\pi N} g_{\rho N}^2 (\mu \kappa_\rho -4) \over 4M^2
(M_\rho^2 +M^2 \mu) (2+\mu)} \Big[ 3 + \mu \Big(2+{\kappa_\rho \over 4} \Big)
\Big] = -0.37 \, {\rm fm}^4\,\,, \\
\label{vmeson3} {\cal B}^{(\omega\rho\pi)}  &=& 0 \,\,. \end{eqnarray}
One observes that the sizeable $\omega$--exchange contribution
${\cal B}^{(\omega)}$ is approximately equal to ${\cal A}^{(\omega)}$ given in
Eq.(\ref{Aom}).  Note also that the expression for $\omega$--exchange cannot be
recovered by simply substituting the vector meson mass and coupling constants
in the expression for $\rho$--exchange. The reason for that are certain
diagrams with (charged) $\rho^+$--exchange which have no analogy in the case of
the (neutral) $\omega$--meson. In the spirit of vector meson dominance one
could also think of an isovector $\pi\rho NN$ contact vertex of the form
\begin{equation} {\cal L}_{\pi \rho N} = {g_\rho g_{\pi N} \over 2M}
\bar N \gamma^\mu \gamma_5 \vec \tau \cdot (\vec \pi \times \vec\rho_\mu\,)N
\,\,. \end{equation}
The form of this vertex and the coupling constant in front are copied from the
Kroll--Ruderman term for charged pion photoproduction replacing the charge $e$
by the universal $\rho$--coupling $g_\rho = 6$ and the photon field by the
isotriplet $\rho$--meson field $\vec \rho_\mu$. The respective
$\rho^0$ and $\rho^+$ exchange diagrams give rise to the following contribution
to the triplet amplitude,
\begin{equation}\label{rhokr} {\cal B}^{(\rho, {\rm KR})} = { g_{\pi N} g_\rho
g_{\rho N} \over M^2 (M_\rho^2 +M^2 \mu) } \Big( 1 - {\mu \over 4} \kappa_\rho
\Big) = 0.45 \, {\rm fm}^4 \,\,. \end{equation}
Admittedly, this contribution is somewhat speculative since the Kroll--Ruderman
vertex for $\rho$--mesons presumes a particular realization of vector meson
dominance. We do not investigate in further detail pion loop diagrams
contributing to ${\cal B}$, since they will turn out to be small in analogy to
the amplitude ${\cal A}$ discussed in section~3.2 (for the classes of
diagrams calculated there). The imaginary part of the triplet amplitude
${\cal B}$ can only be generated in one--pion loop approximation by the
two--pion exchange box diagrams shown in Fig.~9 and evaluated in appendix B.
It can, however, not  be expected that the one--pion loop approximation will be
sufficiently accurate, since the process $NN\to NN\pi$ at threshold is also
sensitive to short distance dynamics. This needs further study but goes beyond
the scope of this paper. A detailed discussion of the imaginary parts of
${\cal A}$ and ${\cal B}$ in one pion loop approximation is given in appendix
B. Summing up the various tree level contributions given in
Eqs.(\ref{isovec},\ref{bpidir},\ref{isoresc},\ref{beta},\ref{vmeson1},\ref{vmeson2},\ref{vmeson3},\ref{rhokr})
we get,
\begin{equation} {\cal B}^{({\rm thy})} = 2.74 \, {\rm fm}^4 \,\,,
\end{equation}
which is very close the real part of the experimental value in Eq.(\ref{Bemp}),
Re$\,{\cal B}^{(\rm exp)}=2.8$ fm$^4$. We thus conclude that also the real part
of the triplet amplitude Re$\,{\cal B}$ can be well understood in terms of
these well-known tree--level meson exchange diagram when evaluated fully
relativistically.  Of course, the relatively large imaginary part Im$\,{\cal
B}^{(\rm exp)} = - 1.5$ fm$^4$ (more than half as large as the
empirical real part) remains unexplained in tree approximation. As mentioned
earlier this imaginary part originates (because of unitarity) from the fact
that the  $^3P_1$ $pp$ phase shift is rather large at the pion production
threshold. The large imaginary part Im$\,{\cal B}$ thus reflects the strong
initial state interaction in the $^3P_1$ entrance channel. For the singlet
transition amplitude ${\cal A}$ the situation is different. Accidentally, the
$^3P_0$ $pp$ phase shift is very small at the pion production threshold and
thus there is only weak initial state interaction in the $^3P_0$ entrance
channel. Another mechanism which could contribute (significantly) to Im$\,{
\cal B}$ is the two--step process $pp\to d \pi^+ \to pn \pi^+$. The threshold
for the deuteron channel $pp \to d \pi^+$ opens 4.8 MeV lower at $T_{\rm lab}=
287.5$ MeV and the corresponding total cross sections are more than an order
of magnitude larger \cite{who,pipn} than the ones for $pp\to pn \pi^+$.

\begin{figure}[t]
   \epsfxsize=10cm
   \centerline{\epsffile{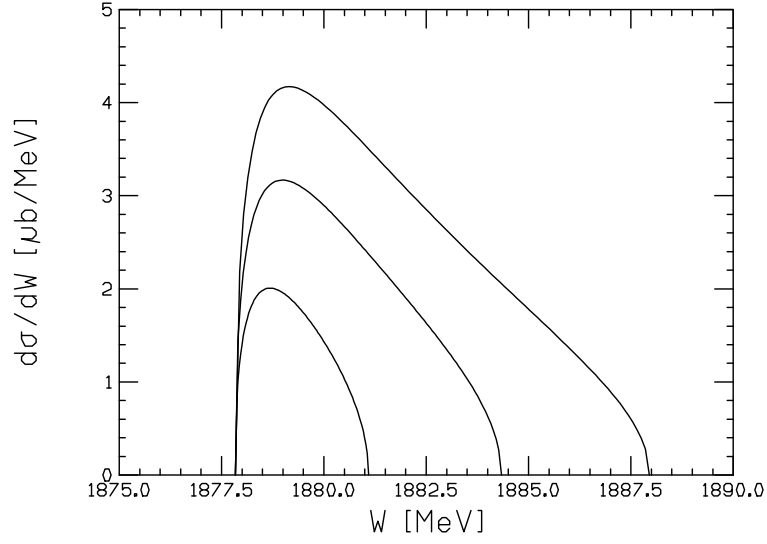}}
   \medskip
   \centerline{\parbox{13cm}{\caption{\label{invmass}
Pion production cross section as a function of the $pn$ invariant mass
$W$. The three curves corresponds to $T_{\rm lab} = 299.3, \, 306.3$
   and $314.1$~MeV, in ascending order. }}}

\end{figure}

We have done several checks on our treatment of the $pn$ final--state
interaction. First, we compared the effective range approximation with the
empirical values of the $^1S_0$ and $^3S_1$ $pn$ phase shifts in the energy
range relevant here and found that deviations are smaller than 2\%. Secondly,
we have studied the $\pi^+$ production cross section as a function of the
$pn$ invariant mass $W$ (given by Eq.(\ref{piplus}) without $dW$--integration)
and found good agreement with the data of \cite{pipn}, cf. Fig.\ref{invmass}.
>From all this we conclude that our treatment of the $pn$ final--state
interaction is fairly realistic.

Finally, we believe that all the features of the processes $pp \to p N\pi$
that we have learned here from the relativistic diagrammatic approach
presented here will be useful for further studies within more complete
dynamical  models, which e.g. treat initial-- and final--state interactions
simultaneously. We also remark that a similar covariant one--boson exchange
model has  been developed in ref.\cite{engel}, which describes the data at
much higher energies, $0.3$ GeV $<T_{\rm lab} <2.0$ GeV. The same
model has been applied to threshold data in ref.\cite{shyam}. These authors
introduce in addition (unobservable) meson nucleon form factors and
energy--dependent coupling constants. This makes a direct comparison between
their work and ours very difficult. However, no isoscalar pion rescattering and
no pion loops are considered. Similar to our finding it is concluded that
$\omega(782)$ exchange is important close to pion production threshold and that
the $\Delta (1232)$ plays no role.

\section{Eta--meson production in $pp$--collisions}

In this section we will discuss $\eta$--production using a similar
approach. The T--matrix for $\eta$--production in proton--proton
collisions, $p_1(\vec p\,) +p_2(-\vec p\,) \to p+p +\eta$, at threshold in the
center--of--mass frame reads (see also appendix A),
\begin{equation}
{\rm T}^{\rm cm}_{\rm th}(pp\to pp \eta) = {\cal C} \, ( i\,\vec
\sigma_1 - i\, \vec \sigma_2+\vec \sigma_1 \times \vec \sigma_2)\cdot \vec p
\,. \end{equation}
with ${\cal C}$ the (complex) threshold amplitude for $\eta$--production. The
$\eta$--production threshold is reached at a proton laboratory kinetic energy
$T_{\rm lab}^{\rm th} = m_\eta(2+m_\eta/2M) = 1254.6$ MeV, where
$m_\eta=547.45$ MeV denotes the eta--meson mass.

\subsection{Extraction of the threshold amplitude}

In the case of $\eta$--production near threshold it is also important to
take into account the $\eta p$ final--state interaction, since the $\eta
N$--system interacts rather strongly near threshold. In fact a recent
coupled--channel analysis~\cite{batinic} of the $(\pi N, \eta N)$-system  finds
for the real part of the $\eta N$ scattering length Re$\,a_{\eta N} =(0.717\pm
0.030)$ fm. For comparison, this value is a factor 5.7 larger than the $\pi^-p$
scattering length, $a_{\pi^- p} = 0.125$ fm \cite{sigg}, measured in pionic
hydrogen.

We assume that the correction due to the S--wave $\eta p$ final--state
interaction near threshold can be treated in effective range approximation
analogous to the S--wave $pp$ final--state interaction. We furthermore make the
assumption that the final state interactions in the $pp$ subsystem and in the
two $\eta p$ subsystems do not influence each other and that they factorize.
These are of course very strong assumptions, but as we will see soon, such a
simple ansatz for the final--state interaction in the $pp\eta$ three--body
system allows to describe rather accurately the energy dependence of the total
cross section $\sigma_{\rm tot}(pp\to pp\eta)$ near threshold. Using the
factorization ansatz mentioned before, the unpolarized total cross section for
$pp\to pp \eta$ reads,
\begin{eqnarray} \label{eta}
\sigma_{\rm tot}(T_{\rm lab}) &=& |{\cal C}|^2
\Big({M \over 4\pi}\Big)^3   {2 \sqrt{T_{\rm lab}} \over(2M+T_{\rm lab})^{3/2}}
\int_{2M}^{W_{\rm max}}d W\,W F_p(W) \nonumber \\
&& \times \int_{s_\eta^-}^{s_\eta^+} ds_\eta
\, F_\eta(s_\eta) \, F_\eta(6M^2+2M T_{\rm lab}+m_\eta^2 -W^2-s_\eta) ~.
\end{eqnarray}
with $W_{\rm max} = \sqrt{4M^2+2M T_{\rm lab}}-m_\eta$ the endpoint of the
di--proton invariant mass spectrum and $F_p(W)$ given by Eq.(6). The variable
$s_\eta$ is the invariant mass squared of the first $\eta p$--pair and the
argument of the last function $F_\eta(\tilde s_\eta)$ in
Eq.(\ref{eta}), $\tilde s_\eta
=6M^2+2M T_{\rm lab}+m_\eta^2 -W^2- s_\eta$, is the invariant mass squared of
the second $\eta p$--pair. The expressions
\begin{equation} \label{bound}
s_\eta^\pm = 3M^2+M T_{\rm lab} +{1\over 2} (m_\eta^2 -W^2)
\pm {1\over 2W} \sqrt{(W^2-4M^2) \lambda(W^2,m_\eta^2, 4M^2+2M T_{\rm
lab})}
\end{equation}
give the boundaries of the $pp\eta$ three--body phase space in the
($s_\eta , W^2$)--plane. Obviously, the formula for the total
cross section, Eq.(\ref{eta}),
is invariant under the permutation of the two $\eta p$--pairs, $s_\eta
\leftrightarrow \tilde s_\eta$, since $s^\pm_\eta = \tilde s^\mp_\eta$.
Furthermore, the correction factor $F_\eta(s_\eta)$ due to the S-wave $\eta p$
final--state  interaction reads in effective range approximation,
\begin{equation} \label{etaeffr}
F_\eta(s_\eta) = \bigg|\, {f^{0+}_{\eta N}(s_\eta) \over a_{\eta N}}\,\bigg|^2
= \bigg|\, 1 - {i\,a_{\eta N}\over 2 \sqrt{s_\eta}}
\sqrt{\lambda(s_\eta, m_\eta^2, M^2)} +{a_{\eta N} \, r_{\eta N}\over 8s_\eta}
\lambda(s_\eta, m_\eta^2, M^2)\,\bigg|^{-2} \,\,. \end{equation}
Here, $f^{0+}_{\eta N}(s_\eta)$ is the S-wave $\eta N$ elastic scattering
amplitude. The (complex) $\eta N$ scattering length $a_{\eta N} = ((0.717 \pm
0.030) + i \, (0.263 \pm 0.025))$ fm is taken from ref.\cite{batinic} and
the (complex) $\eta N$ effective range parameter $r_{\eta N} = ((-1.50\pm 0.13)
- i \,(0.24\pm 0.04))$ fm stems from ref.\cite{green}. It is important to note
that both ref.\cite{batinic} and ref.\cite{green} using quite different methods
agree within error bars on the value of the $\eta N$ scattering length $a_{\eta
N}$.

\begin{figure}[hbt]
   \epsfxsize=10cm
   \centerline{\epsffile{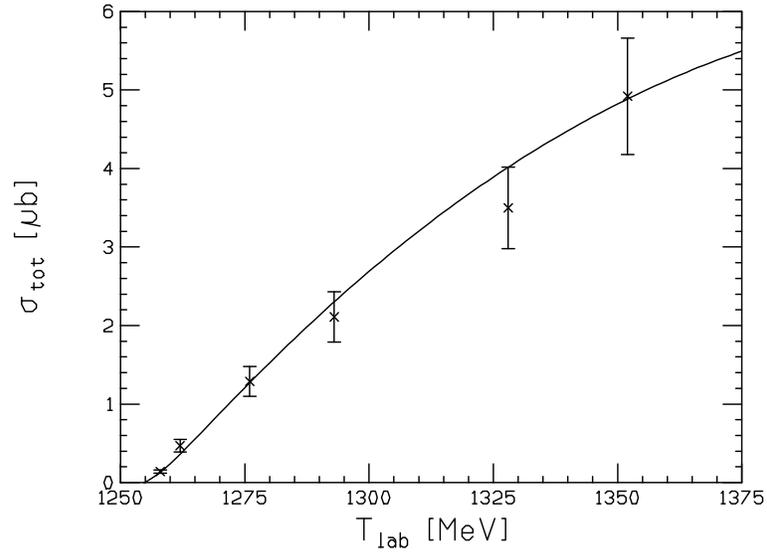}}
   \medskip
   \centerline{\parbox{13cm}{\caption{\label{etaxs}
The eta--production cross section $\sigma_{\rm tot}(pp \to pp \eta)$ as a
function of $T_{\rm lab}$. The  data  are taken from
ref.\cite{calen}.}}}

\end{figure}

Using Eq.(\ref{eta},\ref{bound},\ref{etaeffr}) for the total cross section and
the central values of $a_{\eta N}$ and $r_{\eta N}$ one finds in a least
square fit of the six data points from CELSIUS~\cite{calen} for the modulus of
the threshold amplitude
\begin{equation}
|{\cal C} | = 1.32 \, {\rm fm}^4 \,\,,
\end{equation}
with a total $\chi^2= 3.8$. The resulting energy dependent cross section
from threshold up to $T_{\rm lab} = 1375$ MeV is shown in Fig.~\ref{etaxs}
together with the data from CELSIUS~\cite{calen}.\footnote{Earlier data from
SATURNE \cite{pinot,spes3,hibou} do not have the same accuracy and are not
considered further.} It is rather astonishing that one can describe the total
cross section data up to 100 MeV above threshold with a constant threshold
amplitude ${\cal C}$ and a simple factorization ansatz for the three--body
final--state interaction.

\subsection{Diagrammatic approach}

Next, we turn to the evaluation of the relativistic Feynman diagrams
contributing to $pp\to pp \eta$ at threshold. The resulting expressions can
essentially be copied from the case $pp \to pp \pi^0$ making only the
substitution $(g_{\pi N}, m_\pi) \to (g_{\eta N}, m_\eta)$. One finds for
$\pi^0, \eta, \omega, \rho^0$--exchange
\begin{eqnarray}
{\cal C}^{(\pi^0)} &=&{g_{\eta N} g_{\pi N}^2 m_\eta \over
4M^2( m_\pi^2 +Mm_\eta) (2M+m_\eta)} = 0.17\, {\rm fm}^4 \,\,, \\
{\cal C}^{(\eta,{\rm dir})} &=& {g^3_{\eta N} \over 4M^2(M+
m_\eta)  (2M+m_\eta)} = 0.02\, {\rm fm}^4 \,\,, \\
{\cal C}^{(\omega)} &=& {g_{\eta N} g_{\omega N}^2 (2M-m_\eta)
\over 2M^2( M_\omega^2 +Mm_\eta) (2M+m_\eta)} = 0.23\, {\rm fm}^4 \,\,, \\
{\cal C}^{(\rho^0)} &=& {g_{\eta N} g_{\rho N}^2 \over 2M(
M_\rho^2 +Mm_\eta) (2M+m_\eta)} \nonumber \\ & & \times \bigg[ 2 +{m_\eta
\over M}  (\kappa_\rho^2- \kappa_\rho -1) +{m_\eta^2 \over M^2} \kappa_\rho
\Big(1+{9\over 8} \kappa_\rho \Big) \bigg] = 0.50\, {\rm fm}^4~.
\end{eqnarray}
Note that the $\rho^0$ exchange has become dominant because of the large
tensor--to--vector coupling ratio $\kappa_\rho= 6$ and the larger ratio
$m_\eta/M=0.58$.\footnote{Interestingly, recent measurements of the angular
distributions in $pp\to pp \eta$ suggest the dominance of vector meson exchange
\cite{priv}.}  Besides these diagrams with $\eta$--emission before and
after meson exchange between the protons, one has to account for the strong
$\eta p$ rescattering. Microscopically, the strong $\eta N$ S-wave interaction
originates (among other things) from the nucleon resonance $S_{11}(1535)$ which
is supposed to have a very large coupling to the $\eta N$--channel. Instead of
introducing this resonance together with several parameters (mass, width,
coupling constant), we will merely introduce here a local $NN \eta \eta$
contact vertex of the form
\begin{equation} {\cal L}_{\eta N} = K \, \bar N(x) N(x)\,\eta^2(x) \,\,.
\end{equation}
The interaction strength $K$ is then determined by the real part of the
$\eta N$ scattering length. This means that the pseudovector Born graphs
plus the contact vertex sum up to give the empirical value of Re$\,a_{\eta N}
=0.717\,$fm.  This leads to the equation
\begin{equation}
4\pi \Big( 1 + {m_\eta \over M}\Big)\,{\rm Re\,}a_{\eta N}= 2 K-{g_{\eta
N}^2 m_\eta^2 \over M (4M^2 -m_\eta^2)}  \,\,,
\end{equation}
which results in a value of $K=7.41$ fm. The $\eta$--rescattering graph
(analogous to Fig.~\ref{fig2}c) leads to the following contribution to the
threshold amplitude,
\begin{equation}
{\cal C}^{(\eta,{\rm res})} = {g_{\eta N} \, K \over Mm_\eta
(M+m_\eta) } = 0.40 \,{\rm fm}^4\,\,.
\end{equation}
Evidently, all contributions to ${\cal C}$ scale with the (empirically not well
determined) $\eta N$--coupling constant $g_{\eta N}$. The numbers given in
Eqs.(45,46,47,48,51) which add up to the empirical value of $|{\cal C}|=1.32$
fm$^4$  follow with $g_{\eta N} = 5.3$. Such a value of $g_{\eta N}$ is
consistent with all existing empirical information on it. The SU(3) flavor
symmetry connects  the pion--nucleon and eta--nucleon coupling constants via
the $D/F$ ratio (of the baryon octet axial vector couplings),
\begin{equation} g_{\eta N} = g_{\pi N} { 3-D/F \over \sqrt3 (1+D/F) } \,\,.
\end{equation}
Using $g_{\pi N}=13.4$, the for our purpose optimal value $g_{\eta N}=5.3$
requires a ratio $D/F= 1.37$. In fact a systematic analysis of semileptonic
hyperon decays in ref.\cite{bourquin} gives $D/F = 1.58 \pm 0.07$,
not far from this number. Of course, SU(3) is broken to some extent by the
strange quark mass. For a recent update, see e.g.~\cite{ratc}.
There is one further contribution we have not discussed so far,
related to the $a_0\eta\pi$--coupling. This coupling is rather uncertain.
If one assumes that the diagrams with $\eta$--emission from the
$a_0\eta\pi$--vertex contribute with a positive sign,
\begin{equation}
{\cal C}^{(a_0\eta\pi)} = {g_{a_0 N} g_{\pi N} (c_d m_\eta^2 -2 c_m m_\pi^2)
\over \sqrt{6} M f_\pi^2 (M_{a_0}^2+Mm_\eta)(m_\pi^2 + Mm_\eta)}
= 0.05 \, {\rm fm}^4\,\,,
\end{equation}
one can even lower the $\eta N$--coupling constant to $g_{\eta N}= 5.1$. We
used here $g_{a_0 N}=4.5$~\cite{machleidt} and $c_d= 32\,$MeV,
$c_m = 42\,$MeV~\cite{ecker}. We also note the one--boson exchange model of
\cite{machleidt} for elastic NN--scattering uses an $\eta N$--coupling constant
of $g^{\rm OBE}_{\eta N}=6.8$ not far from our value $g_{\eta N}=5.3$.
The results of our approach applied to the reaction $pn\to pn \eta$ are briefly
discussed in appendix A.

The main point we want to make here is that even the $pp \to pp \eta$
threshold amplitude can be understood in terms of these well--known meson
exchange diagrams when evaluated relativistically. With rather mild assumptions
on the coupling constant $g_{\eta N}$ and the form of the $\eta
N$--rescattering one can easily reproduce the empirical value $|{\cal C}|=
1.32$ fm$^4$. For another boson--exchange approach to $\eta$--production
emphazising the role of the $S_{11}(1535)$ nucleon resonance, see
e.g.~\cite{geda,faeldt} (and references therein).

\subsection{Comments on $\eta '$--production near threshold}

Finally, we like to comment on the recent $\eta'$--production data near
threshold from COSY~\cite{moskal} and SATURNE~\cite{hibou}. Taken
face value, the energy dependence of the four COSY cross section data points
is best described by the pure three--body phase space behavior (as shown by
the dotted curve in Fig.~2 of ref.\cite{moskal}). Of course, it is
hard to imagine
that the $pp$ final-state interaction does not play a role for $pp\to pp
\eta'$ that close to threshold where these data were measured. We have analyzed
the combined COSY and SATURNE data (six data points) within our approach
including the $pp$ final--state interaction in effective range approximation.
Only if one ignores the COSY data at the two lowest energies (at cm excess
energies of 1.5 and 1.7 MeV, respectively),
one can fit the remaining four data points with a modulus of the
threshold amplitude of $|{\cal C}'\,|=0.21\,$fm$^4$ with a small total
$\chi^2 = 2.4$. This fit leads to values of the total cross section at the two
lowest energy points of 5.2 nb and 6.3 nb (see also table~\ref{tab3}). These
numbers  are more than twice as large as the corresponding central values given
in ref.\cite{moskal}. Note, however, that in Fig.2 of
ref.\cite{moskal} sizeable uncertainties in the excess energy are
given, e.g. the lowest point is at $Q = (1.5 \pm 0.5)\,$MeV. Our
discussion always refers to the central $Q$ values.
\begin{table}[hbt]
\begin{center}
\begin{tabular}{|c|cccccc|}
    \hline\hline
    $Q$~[MeV]           & 1.5 & 1.7 & 2.9 & 3.7 & 4.1 & 8.3 \\
    \hline
    $\sigma_{\rm tot}^{\rm exp}$~[nb]
                        & $2.5\pm 0.5$ & $2.9 \pm 1.1$ & $12.7 \pm 3.2$ &
                          $19.2\pm 2.7$ & $25.2 \pm 3.6$ & $43.6 \pm 6.5$ \\
    \hline
    $\sigma_{\rm tot}^{\rm fit}$~[nb]
                        & 5.23 & 6.29 & 13.34 & 18.45 & 21.09 & 50.13 \\
    \hline\hline
  \end{tabular}
\caption{Total cross sections for $pp \to pp\eta '$ as a function of
    the cm excess energy $Q=\sqrt{4M^2+2M T_{\rm lab}} - 2M - M_{\eta '}$
    (only the mean value is given). The data are from
    refs.{\protect{\cite{hibou,moskal}}}. The fit is described in the
    text.}\label{tab3}
\end{center}
\end{table}

Within the relativistic one--boson ($\pi^0,\eta,\omega,\rho^0$) exchange model
the fit value $|{\cal C}'\,| = 0.21$ fm$^4$ implies the relation $g_{\eta' N}
(1-1.28 \,\epsilon) = 1.12$. Here $g_{\eta' N}$ denotes the $\eta' N$--coupling
constant and $\epsilon$ is the fraction of pseudoscalar $\eta' NN$-coupling
(since the $\eta'(958)$ is no Goldstone boson there is no reason to favor the
pseudovector coupling). Interestingly, only the tensor interaction of the
$\rho$--exchange ($\sim \kappa_\rho$) is sensitive to the parameter $\epsilon$.

According to ref.\cite{efremov} one can relate the $\eta' N$--coupling constant
to the quark helicity contribution to the proton spin, $\Delta \Sigma =
\sqrt{3/2} g_{\eta' N} f_\pi /M +0.15$. Using the recent determination $\Delta
\Sigma = 0.45\pm 0.09$ of ref.\cite{altarelli} from  deep inelastic polarized
lepton scattering, one can extract within the relativistic one--boson exchange
model an $\eta' N$--coupling constant of $g_{\eta' N} = 2.5 \pm0.7$ and a
pseudoscalar coupling fraction of $\epsilon = 0.4 \pm0.1$. It remains to be
seen whether other $\eta'$--production processes (e.g. photoproduction $\gamma
p \to \eta' p $) are consistent with these values. The exchange of $\pi^0,
\omega$ and $\rho^0$ contribute to $|{\cal C}'\,|=0.21$ fm$^4$ approximately
$30\%, 20\%$ and $50\%$, respectively.  For another $1\pi$--exchange model to 
$\eta'$--production, see e.g. \cite{sibir}. 

\bigskip

\subsection*{Acknowlegdements}

We thank A. Svarc for information on the $\eta N$ scattering length.
We are particularly grateful to C. Hanhart who pointed out an error concerning
the calculation of ${\cal B}$ in the original version of the manuscript.

\appendix
\def\theequation{\Alph{section}.\arabic{equation}}
\setcounter{equation}{0}

\section{General threshold T--matrices}
In this appendix, we write down the general form of the threshold
T--matrices for $NN\to NN \pi$ and $NN\to NN \eta$ using the isospin formalism
for the two--nucleon system. In the case of (isovector) pion production one
has,
\begin{eqnarray}
{\rm T}^{\rm cm}_{\rm th}(NN\to NN \pi) &=& {{\cal A}\over 2} \, ( i\,\vec
\sigma_1 - i\, \vec \sigma_2+\vec \sigma_1 \times \vec \sigma_2)\cdot \vec p
\,\,\, (\vec \tau_1+\vec \tau_2)\cdot \vec \chi^{\,*} \nonumber \\ & & +
{{\cal B} \over 2} \,(\vec \sigma_1 + \vec \sigma_2)\cdot \vec p  \,\,\, (i\,
\vec \tau_1- i\,\vec \tau_2 + \vec \tau_1 \times \vec \tau_2)\cdot \vec
\chi^{\,*} \,\, .  \end{eqnarray}
with $\vec \chi$ the three--component isospin wave function of the final state
pion, e.g. $\vec \chi = (0,0,1)$ for $\pi^0$--production and $\vec \chi =
(1,i,0)/\sqrt2$ for $\pi^+$--production.  The complex amplitude ${\cal A}$
belongs to the (singlet) transition $^3P_0 \to $ $^1S_0s$ with conserved total
isospin $I_{\rm tot}=1$. Similarly, the complex amplitude ${\cal B}$ belongs to
the (triplet) transition $^3P_1 \to $ $^3S_1s$, also with total isospin $I_{\rm
tot}=1$. In fact the selection rules which follow from the conservation of
parity, angular momentum and isospin allow only for these two transitions. Note
also that there is an invariance under the substitution $({\cal A}, \vec
\sigma_1, \vec \sigma_2, \vec p\,) \leftrightarrow ({\cal B}, \vec
\tau_1, \vec \tau_2, \vec \chi^{\,*})$.

In the case of (isoscalar) eta production one has,
\begin{eqnarray}
{\rm T}^{\rm cm}_{\rm th}(NN\to NN \eta) &=& {{\cal C}\over 4} \, ( i\,\vec
\sigma_1 - i\, \vec \sigma_2+\vec \sigma_1 \times \vec \sigma_2)\cdot \vec p
\,\,\, (3+\vec \tau_1 \cdot\vec \tau_2) \nonumber \\ & & + {{\cal D}\over 4}
\,  (i\,\vec \sigma_1 -i\, \vec \sigma_2- \vec \sigma_1 \times \vec \sigma_2 )
\cdot \vec p  \,\,\, (1-\vec \tau_1 \cdot \vec  \tau_2)\,\, . \end{eqnarray}
Again, the complex amplitude ${\cal C}$ belongs to the (singlet) transition
$^3P_0 \to $ $^1S_0s$ with total isospin $I_{\rm tot}=1$ and the complex
amplitude ${\cal D}$ belongs to the (triplet) transition $^1P_1 \to $ $^3S_1s$
with total isospin $I_{\rm tot}=0$. The determination of the latter amplitude
${\cal D}$ requires measurements of the total cross sections for the process
$pn \to pn \eta$ having neutrons either in the target or the beam. Finally, we
note that the expressions for ${\rm T}^{\rm cm}_{\rm th}(NN \to NN \pi)$ and
${\rm T}^{\rm cm}_{\rm th}(NN \to NN \eta)$ incorporate the Pauli exclusion
principle, since combined left--multiplication  with the spin--exchange
operator $(1+ \vec \sigma_1 \cdot \vec \sigma_2)/2$ and the isospin--exchange
operator $(1+ \vec \tau_1 \cdot \vec \tau_2)/2$ reproduces them up to an
important minus sign.

For the sake of completeness, we give also the contributions of the
one--boson exchange diagrams discussed in section 5.2 to the (triplet)
$\eta$--production amplitude ${\cal D}$. From pseudoscalar meson ($\pi, \eta)$
exchange one finds,
\begin{equation}
 {\cal D}^{(\pi)} = -3 \, {\cal C}^{(\pi^0)} = -0.52\, {\rm fm}^4 \,\,, \qquad
 {\cal D}^{(\eta)} = {\cal C}^{(\eta,{\rm dir})} +{\cal C}^{(\eta,{\rm res})}
 = 0.42\, {\rm fm}^4 \,\,, \end{equation}
(see Eqs.(49,50,55)). In addition vector meson ($\rho, \omega$) exchange gives
rise to the terms,
\begin{eqnarray}
{\cal D}^{(\omega)} &=& {2M+m_\eta \over 2M-m_\eta} \, {\cal C}^{(\omega)}  =
0.41\, {\rm fm}^4 \,\,, \\
{\cal D}^{(\rho)} &=& {3g_{\eta N} g_{\rho N}^2 \over 2M(
M_\rho^2 +Mm_\eta) (2M+m_\eta)} \nonumber \\ & & \times \bigg[ -2 +{m_\eta
\over M}  (\kappa_\rho^2+ \kappa_\rho -1) +{m_\eta^2 \over M^2} \kappa_\rho
\Big(1+{7\over 8} \kappa_\rho \Big) \bigg] = 1.50\, {\rm fm}^4\,\,,
\end{eqnarray}
(see Eq.(51)). These contributions sum up to ${\cal D}^{(\rm thy)}=
1.81$ fm$^4$. The expression for the total cross section $\sigma_{\rm tot}
(T_{\rm lab})$ of the reaction $pn\to pn \eta$ is obtained from Eq.(45)
replacing the factor $2|{\cal C}|^2 F_p(W)$ by $|{\cal C}|^2 F_s(W)+|{\cal
D}|^2 F_t(W)$ (see Eq.(31)) and we have neglected the small neutron--proton
mass difference.

Recently, total cross sections for $pn\to pn \eta$ have been extracted from
measurements of the process $pd\to ppn\eta$ at CELSIUS \cite{haegg}. A fit of
the two data points closest to threshold (at $T_{\rm lab}=1296$ and $1322$ MeV)
gives for the empirical triplet $\eta$--production amplitude $|{\cal D}|^{(\rm
exp)} =2.3$ fm$^4$. Compared to this value the abovementioned theoretical
prediction, ${\cal D}^{(\rm thy)} = 1.8$ fm$^4$, is only about $20\%$ too
small.

\def\theequation{\Alph{section}.\arabic{equation}}
\setcounter{equation}{0}
\section{Imaginary parts from one--pion loop graphs}

In this appendix we will give explicit expressions for the imaginary parts of
the pion production threshold amplitudes Im$\,{\cal A}$ and Im$\,{\cal B}$ as
they arise in one--pion loop approximation. To that order any non--vanishing
imaginary part can only come from those one--loop diagrams which involve proton
rescattering in the initial state. These are just the two--pion exchange box
diagrams shown in Fig.~9. We apply the Cutkosky cutting rules to evaluate
their imaginary part. It is then given by the product of the upper tree--level
subgraph (i.e. the threshold pion production amplitude with $1\pi$--exchange
including the leading order isovector pion rescattering) and of one--half the
invariant $pp$ two-body phase space times the lower tree--level subgraph. The
latter two factors combine to the $^3P_0$ or $^3P_1$ $pp$  phase shift
calculated perturbatively in $1\pi$--exchange approximation at
$T_{\rm lab}^{\rm th} = m_\pi(2+m_\pi/2M)$. Altogether one gets thus from
perturbative unitarity,
\begin{equation} {\rm Im}\,{\cal A}= {\cal A}^{(\pi, {\rm dir})} \cdot
\delta_{1\pi}(^3P_0) \,\,, \end{equation}
\begin{equation} {\rm Im}\,{\cal B}= \Big({\cal B}^{(\pi, {\rm iv})}_{|c_4=0}
+ {\cal B}^{(\pi, {\rm dir})}  \Big)\cdot\delta_{1\pi}(^3P_1)\,\,,
\end{equation}
with ${\cal A}^{(\pi,{\rm dir})}$, ${\cal B}^{(\pi,\rm iv)}_{|c_4=0}$ and
${\cal B}^{(\pi, \rm dir)}$ given in
Eqs.(\ref{pidir},\ref{isovec},\ref{bpidir}). Using the projection formulas of
ref.\cite{nnpap} to calculate the $1\pi$--exchange $^3P_{0,1}$ phase shifts one
finds the following analytical results,
\begin{equation} {\rm Im}\,{\cal A} ={g_{\pi N}^5 \sqrt{\mu(4+\mu)} \over 64
\pi M^4 (1+\mu)(2+\mu)^2 } \bigg[ 1 - {\mu \over 4+\mu} \ln\Big( 2+{4\over \mu}
\Big) \bigg] = 0.54\, {\rm fm}^4 \,\,, \end{equation}
\begin{eqnarray} {\rm Im}\,{\cal B} &=&{g_{\pi N}^5 \sqrt{\mu(4+\mu)}\over 64
\pi M^4 (1+\mu)(2+\mu)^2 }\Big(3+2 \mu -2g_A^{-2}(2+\mu)\Big)\\ \nonumber & &
\times \bigg[ {\mu-4 \over 2(4+\mu)} - {\mu^2 \over (4+\mu)^2}  \ln\Big(
2+{4\over \mu} \Big) \bigg] = -0.24\,  {\rm fm}^4 \,\,,\end{eqnarray}
with $g_A=g_{\pi N} f_\pi/M$.
One observes that in one--pion loop approximation Im$\,{\cal A}$ is too large
by a factor of 2 with the wrong (positive) sign, whereas Im$\,{\cal B}$ is to
small by a factor of 6. Clearly, there is important short range NN--dynamics
missing in one--pion loop approximation as can be seen by comparing the
empirical $^3P_{0,1}$ phase shifts at $T_{\rm lab}^{\rm th}=m_\pi(2+m_\pi/2M)$,
$\delta(^3P_0)=-6.3^\circ$ and $\delta(^3P_1)=-28.1^\circ$, with the
$1\pi$--exchange approximation, $\delta_{1\pi}(^3P_0)= +65.0^\circ$ and
$\delta_{1\pi}(^3P_1) = -34.6^\circ$. If one considers all those one--loop box
diagrams which include in the upper part all tree graphs evaluated in section
4.2 and in the lower part only the one--pion exchange one gets for the
imaginary part of ${\cal B}$,
\begin{equation} {\rm Im}\, {\cal B}= {\cal B}^{(\rm thy)} \cdot \delta_{1\pi}
(^3P_1)= 2.74\, {\rm fm}^4 \cdot (-0.604) = -1.65 \, {\rm fm}^4 \,\,,
\end{equation}
a value which is in $10\%$ agreement with the empirical Im$\,{\cal B}=-1.5$
fm$^4$. Of course, this merely reflects the fact that the empirical $^3P_1$
phase shift at $T_{\rm lab}^{\rm th}$ is not far from one obtained in
$1\pi$--exchange approximation. Note, however, that for the $^3P_0$ phase
shift at $T_{\rm lab}^{\rm th}$ the empirical value and the $1\pi$--exchange
approximation differ by a large factor of $-10$.

\def\theequation{\Alph{section}.\arabic{equation}}
\setcounter{equation}{0}
\section{Effective field theory approach to final--state interaction}
In this appendix we want to give an elementary derivation of the final-state
interaction correction factor Eq.(6), $F_p(W) = [1+a_p^2 P_*^2]^{-1}, \,P_*^2= 
W^2/4 -M^2$, in scattering length approximation (i.e. for $r_p=0$). Close to 
threshold all final state three--momenta are small and therefore one can
approximate both the meson production process $NN\to NN\pi^0$ and elastic 
$NN\to NN$ scattering by momentum independent contact vertices proportional to
$\cal A$ and the scattering length $a_p$, respectively. Consider first low
energy $NN$--scattering in this approximation. The bubble diagrams with
0,1,2,\dots rescatterings can be easily summed up in the form of a geometric 
series,
\begin{equation} a_p - 4 \pi a_p^2 \int{d^3l\over (2\pi)^3} {1\over
P_*^2+i0^+-l^2} + \dots = a_p + i\, a_p^2 P_* + \dots = {a_p \over 1 - i\, a_p
P_*} \,\,, \end{equation} 
using dimensional regularization to evaluate the (vanishing) real part of the
loop integral. Obviously, the sum of these infinitely many loop diagrams is
just the unitarized scattering length approximation which leads to, 
\begin{equation} \tan \delta_0(W) = a_p P_*\,\,. \end{equation}            
Next, consider in the same approximation meson production followed by an
arbitrary number of  $NN$--rescatterings in the final--state.  Again, these
loop diagrams can be summed up to, 
\begin{equation} {{\cal A} \over 1 - i \, a_p P_*}\,\,, \end{equation}
and taking the absolute square, 
\begin{equation} \bigg|{{\cal A} \over 1 - i\, a_p P_*} \bigg|^2 = {|{\cal
A}|^2 \over 1 + a_p^2 P_*^2} \,\,, \end{equation} 
one encounters the final--state interaction correction factor $F_p(W)=[1+a_p^2
(W^2/4-M^2)]^{-1}$ in scattering length approximation. Since, the scattering
length is much bigger than the effective range parameter for $NN$--scattering
($a_p >> r_p$) one has already derived the dominant effect due to final--state
interaction.  Of course, in order to be more accurate one should eventually go
beyond momentum independent contact vertices. The main point we want to make
here is that the final--state interaction correction factor $F_p(W)$ (for
$r_p=0$) has a sound foundation in effective field theory.


\bigskip

\section*{Figures}

\vspace{1.6cm}

\vspace{1.6cm}

\begin{figure}[t]
   \epsfxsize=10cm
   \centerline{\epsffile{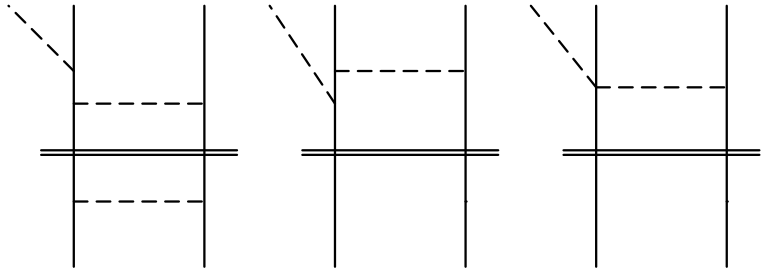}}
   \medskip
    \centerline{\parbox{13cm}{\caption{\label{imagin}
Diagrams that give rise to the non--vanishing imaginary parts Im$\,{\cal A}$
and Im$\,{\cal B}$ in one--pion loop approximation. The projection on the
on--shell $pp$ intermediate state is symbolized by the double
line cutting the diagrams. For further notation, see Fig.~3. }}}

\end{figure}
\end{document}